\documentclass[11pt,twoside]{article}

\usepackage{amsmath,amsfonts,amssymb,amsthm}
\usepackage{epsfig}
\newcommand{\rd}{{\rm d}}

\newcommand{\field}[1]{\mathbb{#1}}
\newcommand{\N}{\field{N}}
\newcommand{\R}{\field{R}}
\newcommand{\C}{\field{C}}

\newcommand{\F}{\mathcal{F}}
\renewcommand{\H}{\mathcal{H}}

\newcommand{\dGamma}{\mathrm{d}\Gamma}
\newcommand{\ad}[2]{\mathrm{ad}_{#1}(#2)}
\newcommand{\add}[2]{\mathrm{ad}_{#1}^2(#2)}

\newcommand{\eps}{\varepsilon}
\newcommand{\ph}{\varphi}
\newcommand{\const}{\mathrm{const}}
\newcommand{\ran}{\mathrm{Ran}}

\newcommand{\egap}{e_{\mathrm{gap}}}

\newcommand{\Hpart}{H_{\text{part}}}
\newcommand{\Ppart}{P_{\text{part}}}

\newcommand{\expect}[1]{\mbox{$\langle #1 \rangle $}}         % expectation value
\newcommand{\sprod}[2]{\mbox{$\langle #1,#2 \rangle$}}       % scalar product
\newcommand{\supp}{\operatorname{supp}}
\newcommand{\Ima}{\operatorname{Im}}
\newcommand{\Rea}{\operatorname{Re}}

\renewcommand{\div}{\mathrm{div}}

\newtheorem{theorem}{Theorem}
\newtheorem{lemma}[theorem]{Lemma}

\newtheorem{prop}[theorem]{Proposition}

\begin{document}
\title{\bf Spectral Theory for the Standard Model
of Non-Relativistic QED}\footnotetext [3]{Supported by NSERC under Grant NA 7901}

\author{\vspace{5pt} J. Fr\"ohlich$^1$\footnote{juerg@itp.phys.ethz.ch} ,
M. Griesemer$^2$\footnote{marcel@mathematik.uni-stuttgart.de}
and I.M.~Sigal$^{3}$\footnote{im.sigal@utoronto.ca} \\
\vspace{-4pt}\small{$1.$ Theoretical Physics, ETH--H\"onggerberg,} \\
\small{CH--8093 Z\"urich, Switzerland}\\
\vspace{-4pt}\small{$2.$ Department of Mathematics, University of
Stuttgart,} \\
\small{D--70569 Stuttgart, Germany}\\
\vspace{-4pt}\small{$3.$ Department of Mathematics, University of
Toronto,} \\
\small{Toronto, Ontario, Canada M5S 2E4}\\ }

\date{\small 28 August, 2007}

\maketitle
\begin{abstract}
For a model of atoms and molecules made from static nuclei and
non-relativistic electrons coupled to the quantized radiation
field (the standard model of non-relativistic QED), we prove a
Mourre estimate and a limiting absorption principle in a
neighborhood of the ground state energy. As corollaries we derive
local decay estimates for the photon dynamics, and we prove
absence of (excited) eigenvalues and absolute continuity of the
energy spectrum near the ground state energy, a region of the
spectrum not understood in previous investigations.

The conjugate operator in our Mourre estimate is the second
quantized generator of dilatations on Fock space.
\end{abstract}

\section{Introduction}
According to Bohr's well known picture, an atom or molecule has
only a discrete set of stationary states (bound states) at low
energies and a continuum of states at energies above the
ionization threshold. Electrons can jump from a stationary state
to another such state at lower energy by emitting photons. These
radiative transitions tend to render excited states unstable,
i.e., convert them into resonances. Exceptions are the
\emph{ground state} and, in some cases, excited states that remain
stable for reasons of symmetry (e.g. ortho-helium). In
\emph{non-relativistic QED}, the instability of excited states
finds its mathematical expression in the migration of eigenvalues
to the lower complex half-plane (second Riemannian sheet for a
weighted resolvent) as the interaction between electrons and
photons is turned on. Indeed, the spectrum of the Hamiltonian
becomes purely \emph{absolutely continuous} in a neighborhood of
the unperturbed excited eigenvalues \cite{BFSS, BFS3}. The ground
state, however, remains stable \cite{BFS1, BFS3, GLL}. The methods
used to analyze the spectrum near unperturbed excited eigenvalues
have either failed \cite{BFSS}, or not been pushed far enough
\cite{BFS3}, to yield information on the nature of the spectrum of the
interacting Hamiltonian in a neighborhood of the ground state
energy. The purpose of this paper is to close this gap: we
establish a Mourre estimate and a corresponding limiting
absorption principle for a spectral interval at the infimum of the
energy spectrum. It follows that the spectrum is purely absolutely
continuous above the ground state energy. As a corollary we prove
local decay estimates for the photon dynamics.

In non-relativistic QED (regularized in the ultraviolet), the
Hamiltonian, $H$, of an atom or molecule with static nuclei with
is a self-adjoint operator on the tensor product, $\H:=\H_{\rm
part}\otimes \F$, of the electronic Hilbert space $\H_{\rm part} =
\wedge_{i=1}^N L^2(\R^3;\C^2)$ and the symmetric (bosonic) Fock
space $\F$ over $L^2(\R^3,\C^2;dk)$. It is given by
\begin{equation}\label{intro:Ham}
   H= \sum_{i=1}^{N} (-i\nabla_{x_i}+\alpha^{3/2} A(\alpha x_i))^2 + V +
   H_{f},
\end{equation}
where $N$ is the number of electrons and $\alpha>0$ is the fine
structure constant. The variable $x_i\in\R^3$ denotes the position
of the $i$th electron, and $V$ is the operator of multiplication
by $V(x_1,\ldots,x_N)$, the potential energy due to the
interaction of the electrons and the nuclei through their
electrostatic fields. In our units, $V(x_1,\ldots,x_N)$ is
independent of $\alpha$ and given by
\begin{equation*}
  V(x_1,\ldots,x_N) = -\sum_{i=1}^N\sum_{l=1}^{M}\frac{Z_l}{|x_i-R_l|} +
  \sum_{i<j}\frac{1}{|x_i-x_j|}.
\end{equation*}
The operator $H_f$ accounts for the energy of the transversal
modes of the electromagnetic field, and $A(x)$ is the quantized
vector potential in the Coulomb gauge with an ultraviolet cutoff.
In terms of creation- and annihilation operators,
$a^{*}_{\lambda}(k)$ and $a_{\lambda}(k)$, these operators are
\begin{equation*}
  H_{f}=\sum_{\lambda=1,2}\int d^3k |k|
a^{*}_{\lambda}(k)a_{\lambda}(k),
\end{equation*}
and
\begin{equation}\label{intro:A-pot}
   A(x) = \sum_{\lambda=1,2}\int d^3k \frac{\kappa(k)}{|k|^{1/2}}
   \eps_{\lambda}(k)\Big\{e^{ik\cdot x} a_{\lambda}(k) + e^{-ik\cdot x}
   a^{*}_{\lambda}(k)\Big\},
\end{equation}
where $\lambda\in \{1,2\}$ labels the two possible photon
polarizations perpendicular to $k\in \R^3$. The corresponding
polarization vectors are denoted by $ \eps_{\lambda}(k)$; they are
normalized and orthogonal to each other. Thus, for each $x\in
\R^3$, $A(x)=(A_1(x),A_2(x),A_3(x))$ is a triple of operators on the Fock
space $\F$. The real-valued function $\kappa$ is an ultraviolet
cutoff and serves to make the components of $A(x)$ densely defined
self-adjoint operators. We assume that
$\kappa$ belongs to the Schwartz space, although much less
smoothness and decay suffice. We emphasize that no infrared
cutoff is used; that is, (physically relevant) choices of
$\kappa$, with
\begin{equation}\label{kappa0}
   \kappa(0)\neq 0
\end{equation}
are allowed. The spectral analysis of $H$ for such choices of
$\kappa$ is the main concern of this paper. Under the simplifying
assumption that $|\kappa(k)|\leq |k|^{\beta}$, for some $\beta>0$,
the analysis is easier and some of our results are already known
for $\beta$ sufficiently large; see the brief review at the end of this introduction.

The spectrum of $H$ is the half-line $[E,\infty)$, with
$E=\inf\sigma(H)$. The end point $E$ is an eigenvalue if
$N-1<\sum_{l}Z_j$ \cite{BFS2, GLL, LiebLoss2003}, but the rest of
the spectrum is expected to be purely absolutely continuous (with
possible exception as explained above). For a large interval
between $E$ and the threshold, $\Sigma$, of ionization, absolute
continuity has been proven in \cite{BFSS, BFS2}; but the nature of
the spectrum in small neighborhoods of $E$ and $\Sigma$ has remained
open. There are further results on absolute continuity of
the spectrum for simplified variants of $H$, and we shall comment
on them below. 

Our first main result concerns the spectrum of $H$
in a neighborhood of $E$. Under the assumptions that $\alpha$ is sufficiently
small and that $e_1=\inf\sigma(\Hpart)$ is a simple and isolated eigenvalue
of $\Hpart=-\sum_{i=1}^N\Delta_{x_i}+V$, we show that $\sigma(H)$
is purely absolutely continuous in $(E,E+\egap/3)$, where $\egap
=e_2-e_1$ and $e_2$ is the first point in the spectrum of $\Hpart$
above $e_1$. It follows, in particular, that $H$ has no
eigenvalues near $E$ other than $E$. Our second main result concerns the
dynamics of states in the spectral subspace of $H$ associated with the
interval $(E,E+\egap/3)$. If $f\in \C_0^{\infty}(\R)$ with 
$\supp(f)\subset (E,E+\egap/3)$, then
\begin{equation}\label{locdecay}
    \|\expect{B}^{-s}e^{-iHt}f(H)\expect{B}^{-s}\| =
    O(\frac{1}{t^{s-1/2}}),\qquad (t\to \infty),
\end{equation}
where $B$, is the second quantized dilatation generator on Fock space, that is,
\begin{equation}\label{intro0}
    B=\dGamma(b),\qquad b=\frac{1}{2}(k\cdot y + y\cdot k).
\end{equation}
Here $y=i\nabla_k$ denotes the ``position operator'' for photons and $\expect{B}:=(1+B^2)^{1/2}$. 
Estimate \eqref{locdecay} is a statement about the growth of
$B$ under the time evolution of states in the range of $f(H)\expect{B}^{-s}$. 
Since growth of $B$ requires that either the number of photons or their
distance to the atom grows, \eqref{locdecay} confirms the expectation that,
asymptotically as time tends to $\infty$, the state of an 
excited atoms or molecule relaxes to the ground state by emission of
photons, provided the maximal energy is below the ionization threshold
\cite{Spohn, FGS1, Gerard2002}. In the course of this process the atom or molecule (not including the photons that were radiated
off) will eventually wind up, energetically, in a neighborhood of the ground
state energy $E$. Hence the importance of understanding the spectrum of $H$
and the dynamics generated by $H$ in spectral subspaces of energies near $E$.
We remark that the details of the form of interaction between matter and radiation as given in \eqref{intro:Ham} and
\eqref{intro:A-pot} are \emph{essential} for our results to hold, but that our methods are applicable to other
models of matter and radiation as well, and our corresponding results will be
published elsewhere.

Our approach to the spectral analysis of $H$ is based on Conjugate
Operator Theory in its standard form with a \emph{self-adjoint}
conjugate operator. Our choice for the conjugate operator is
the second quantized dilatation generator \eqref{intro0}. 
The hypotheses of conjugate operator theory are a regularity
assumption on $H$ and a positive commutator estimate, called
\emph{Mourre estimate}. Concerning the first assumption we show
that $s\mapsto e^{-iBs}f(H)e^{iBs}\psi $ is twice continuously
differentiable, for all $\psi\in\H$ and for all $f$ of class
$C_0^{\infty}$ on the interval $(-\infty,\Sigma)$ below the
ionization threshold $\Sigma$. Our Mourre estimate says that, if $\alpha$ is
small enough, then
\begin{equation}\label{intro1}
  E_{\Delta}(H-E)[H,iB]E_{\Delta}(H-E) \geq
  \frac{\sigma}{10}E_{\Delta}(H-E),
\end{equation}
for arbitrary $\sigma\leq \egap/2$ and $\Delta=[\sigma/3,2\sigma/3]$.
As a result we obtain all the standard consequences of conjugate
operator theory on the interval $(E,E+\egap/2)$
\cite{Sahbani1997}, in particular, absence of eigenvalues (Virial
Theorem), absolute continuity of the spectrum, existence of the
boundary values
\begin{equation}\label{intro2}
   \expect{B}^{-s}(H-\lambda\pm i0)^{-1}\expect{B}^{-s}
\end{equation}
for $\lambda \in (E,E+\egap/3)$, $s\in(1/2,1)$ (Limiting
Absorption Principle), and their H\"older continuity of degree
$s-1/2$ with respect to $\lambda$. This H\"older continuity
implies the local decay estimate \eqref{locdecay}.

The idea to use conjugate operator theory with \eqref{intro0} as
the conjugate operator is not new and has been used for instance
in \cite{BFSS}. It is based on the property that
$$
     [H_f,iB] = H_f
$$
and that $H_f$ is positive on the orthogonal complement of the
vacuum sector. There is an obvious problem, however, with the
implementation of this idea that discouraged people from using it
in the analysis of the spectrum close to $E$: if
$\alpha^{3/2}W=H-(H_{\rm part}+H_f)$ denotes the interaction part
of $H$, then
\begin{equation}\label{intro3}
  [H,iB] = H_f + \alpha^{3/2}[W,iB],
\end{equation}
and the commutator $[W,iB]$ has \emph{no definite sign}. It can be
compensated for by part of the field energy $H_f$ so that $H_f +
\alpha^{3/2}[W,iB]$ becomes positive, but only so on spectral
subspaces corresponding to energy intervals separated from $E$ by a distance of
order $\alpha^3$ \cite{BFSS}. For fixed $\alpha>0$ no positive
commutator, and thus no information on the spectrum is obtained
near $E=\inf\sigma(H)$. For this reason, H\"ubner and Spohn
and, later, Skibsted, Derezi\'nski and Jak\v{s}i\'c, and Georgescu
et al. chose the operator
\[
   \hat{B} = \frac{1}{2}\dGamma(\hat{k}\cdot y+ y\cdot
   \hat{k}),\qquad \hat{k}=\frac{k}{|k|},
\]
or a variant thereof, as conjugate operator; see \cite{HubnerSpohn1995a, Skibsted1998,
  DerezinskiJaksic2001, GGM2}. It has the advantage that, formally,
$[H_f,i\hat{B}]=N$, the number operator, which is bounded below by
the identity operator on the orthogonal complement of the vacuum
sector. It follows that $[H,i\hat{B}]\geq \frac{1}{2}N$, for
$\alpha>0$ small enough, and one may hope to prove absolute
continuity of the energy spectrum all the way down to
$\inf\sigma(H)$. The drawback of $\hat{B}$ is that it is only
symmetric, but not self-adjoint, and hence not admissible as a
conjugate operator. Therefore Skibsted, and, later, Georgescu,
G\'erard, and M{\o}ller developed suitable extensions of conjugate
operator theory that allow for non-selfadjoint conjugate operators
\cite{Skibsted1998, GGM2}. Skibsted applied his conjugate operator
theory to \eqref{intro:Ham} and obtained absolute continuity of
the energy spectrum away from thresholds and eigenvalues under an
\emph{infrared} (IR) regularization, but not for \eqref{kappa0}.
For the spectral results of Georgescu et al.~see the review below.
Given this background, the \emph{main achievement} of the present
paper is the discovery of the Mourre estimate \eqref{intro1}. We
now sketch the main elements of its proof.

1. As an auxiliary operator we introduce an IR-cutoff Hamiltonian
$H_{\sigma}$ in which the interaction of electrons with photons
of energy $\omega\leq \sigma$ is turned off. It follows that
$H_{\sigma}$ is of the form
\[
   H_{\sigma} = H^{\sigma}\otimes 1 + 1\otimes H_{f,\sigma},
\]
with respect to $\H = \H^{\sigma}\otimes \F_{\sigma}$, where
$\F_{\sigma}$ is the symmetric Fock space over $L^2(|k|\leq
\sigma;\C^2)$ and $H_{f,\sigma}$ is $\dGamma(\omega)$ restricted
to $\F_{\sigma}$. We show that the reduced Hamiltonian
$H^{\sigma}$ does not have spectrum in the interval $(E_{\sigma},E_{\sigma}+\sigma)$
above the ground state energy
$E_{\sigma}=\inf\sigma(H_{\sigma})=\inf\sigma(H^{\sigma})$. It
follows that, for any $\Delta\subset(0,\sigma)$,
\begin{equation}\label{intro1a}
  E_{\Delta}(H_{\sigma}-E_{\sigma}) = P^{\sigma}\otimes
   E_{\Delta}(H_{f,\sigma}),
\end{equation}
where $P^{\sigma}$ is the ground state projection of
$H^{\sigma}$.

2. We split $B$ into two pieces $B=B_{\sigma}+B^{\sigma}$ where
$B_{\sigma}$ and $B^{\sigma}$ are the second quantizations of the
generators associated with the vector fields $\eta_{\sigma}^2(k)k$
and $\eta^{\sigma}(k)^2k$, respectively. Here
$\eta_{\sigma},\eta^{\sigma}\in C^{\infty}(\R^3)$ is a partition
of unity, $\eta_{\sigma}^2+(\eta^{\sigma})^2=1$, with
$\eta_{\sigma}(k)=1$ for $|k|\leq 2\sigma$ and
$\eta^{\sigma}(k)=1$ for $|k|\geq 4\sigma$. It follows that
$B^{\sigma}=B^{\sigma}\otimes 1$ with respect to $\H =
\H^{\sigma}\otimes \F_{\sigma}$, and that
$[H,B^{\sigma}]=[H^{\sigma},B^{\sigma}]\otimes 1$. Thus
\eqref{intro1a} and the virial theorem,
$P^{\sigma}[H^{\sigma},B^{\sigma}]P^{\sigma}=0$, imply that
\begin{equation}\label{intro2a}
  E_{\Delta}(H_{\sigma}-E_{\sigma})[H,iB^{\sigma}]E_{\Delta}(H_{\sigma}-E_{\sigma}) =0.
\end{equation}

3. The first key estimate in our proof of \eqref{intro1} is the
operator inequality
\begin{equation}\label{intro3a}
   E_{\Delta}(H_{\sigma}-E_{\sigma})[H,iB_{\sigma}]E_{\Delta}(H_{\sigma}-E_{\sigma})
   \geq \frac{\sigma}{8} E_{\Delta}(H_{\sigma}-E_{\sigma})
\end{equation}
valid for the interval $\Delta=[\sigma/3,2\sigma/3]$ and $\alpha\ll 1$, with
$\alpha$ independent of $\sigma$. This
inequality follows from
\begin{equation}\label{intro3b}
  [H_f,iB_{\sigma}] = \dGamma(\eta_{\sigma}^2\omega) \geq H_{f,\sigma}
\end{equation}
and from
\begin{equation}\label{intro3c}
    E_{\Delta}(H_{\sigma}-E_{\sigma})[\alpha^{3/2}H_f+\alpha^{3/2}W,iB_{\sigma}]
    E_{\Delta}(H_{\sigma}-E_{\sigma}) \geq O(\alpha^{3/2}\sigma).
\end{equation}
Indeed, by writing $H_f=(1-\alpha^{3/2})H_f + \alpha^{3/2}H_f$,
combining \eqref{intro3b} and \eqref{intro3c}, and using \eqref{intro1a} we obtain
\begin{equation}\label{intro3d}
   E_{\Delta}(H_{\sigma}-E_{\sigma})[H,iB_{\sigma}]E_{\Delta}(H_{\sigma}-E_{\sigma})
   \geq \big((1-\alpha^{3/2})\inf\Delta+O(\alpha^{3/2}\sigma)\big)
   E_{\Delta}(H_{\sigma}-E_{\sigma}).
\end{equation}
For $\Delta=[\sigma/3,2\sigma/3]$ and $\alpha$ small enough this
proves \eqref{intro3a}.

4. The second key estimate in our proof of \eqref{intro1} is the
norm bound
\begin{equation}\label{intro4a}
  \|f_{\Delta}(H-E)-f_{\Delta}(H_{\sigma}-E_{\sigma})\| = O(\alpha^{3/2}\sigma)
\end{equation}
valid for smoothed characteristic functions $f_{\Delta}$ of the
interval $\Delta=[\sigma/3,2\sigma/3]$. The Mourre estimate
\eqref{intro1} follows from \eqref{intro2a}, \eqref{intro3a}, from
$B=B_{\sigma}+B^{\sigma}$ and from \eqref{intro4a} if $\alpha\ll
1$, with $\alpha$ independent of $\sigma$.

We conclude this introduction with a review of previous work
closely related to this paper. Absolute continuity of (part of)
the spectrum of Hamiltonians of the form \eqref{intro:Ham}, or
caricatures thereof, was previously established in
\cite{HubnerSpohn1995a, Arai1983, Skibsted1998,
GGM2, BFS1, BFS2, BFSS}. Arai considers the explicitly solvable
case of a harmonically bound particle coupled to the quantized
radiation field in the dipole approximation. H\"ubner and Spohn
study the spin-boson model with massive bosons or with a photon
number cutoff imposed. Their work inspired \cite{Skibsted1998}
and \cite{GGM2}, where better results were obtained: Skibsted
analyzed \eqref{intro:Ham} and assumed that $|\kappa(k)|\leq
|k|^{5/2}$, while, in \cite{GGM2}, $|\kappa(k)|\leq |k|^{\beta}$,
with $\beta>1/2$, is sufficient for a Nelson-type model with
scalar bosons. The main achievement of \cite{GGM2} is that no bound on
the coupling strength is required. Papers \cite{BFS2} and
\cite{BFSS} do not introduce an infrared regularization but
establish the spectral properties mentioned above only away from
$O(\alpha^3)$-neighborhoods of the particle ground state energy
and the ionization threshold.

%%%%%%%%%%%%%%%%%%%%%%%%%%%%%%%%%%%%%%%%%%%%%%%%%%%%%%%%%%%%%%%%%%%%%%%%%%%%%%%%%%%%%%%%%

\section{Notations and Main Results}
\label{sec:vector}

This section describes in detail the class of Hamiltonians to which we shall 
apply our analysis, and it contains all our main results. For clarity and
simplicity of the presentation of our techniques and main ideas, we shall restrict 
ourselves to a one-electron model where spin is neglected. Our analysis can easily be
extended to the many electron model presented in the introduction, and spin
may be included as well. 

The Hilbert space of our systems is the tensor product
$$
     \H = L^2(\R^3,dx)\otimes \F,
$$
where $\F$ denotes the symmetric Fock space over $L^2(\R^3;\C^2)$.
The Hamiltonian $H:D(H)\subset\H\to\H$ is given by
\begin{equation}\label{vector-ham}
  H= \Pi^2 + V+ H_f,\qquad \Pi=-i\nabla_x + \alpha^{3/2}A(\alpha x)
\end{equation}
where $V$ denotes multiplication with
a real-valued function $V\in L^2_{\rm loc}(\R^3)$. We assume that
$V$ is $\Delta-$bounded with relative bound zero and that $e_1=\inf\sigma(-\Delta+V)$ is an isolated
eigenvalue with multiplicity one. The first point in
$\sigma(-\Delta+V)$ above $e_1$ is denoted by $e_2$ and
$\egap:=e_2-e_1$. The field energy $H_f$ and the quantized vector
potential have already been introduced, formally, in the
introduction. More proper definitions are $H_f:=\dGamma(\omega)$,
the second quantization of multiplication with $\omega(k)=|k|$,
and $A_j(\alpha x) = a(G_{x,j})+a^{*}(G_{x,j})$ where
$$
     G_{x}(k,\lambda) :=
     \frac{\kappa(k)}{\sqrt{|k|}}\eps_{\lambda}(k)e^{-i\alpha x\cdot
     k},
$$
and $\eps_{\lambda}(k)$, $\lambda\in\{1,2\}$, are two polarization
vectors that, for each $k\neq 0$, are perpendicular to $k$ and to
one another. We assume that $\eps_{\lambda}(k)=\eps_{\lambda}(k/|k|)$. 
The ultraviolet cutoff $\kappa:\R^3\to \C$ is assumed to be a Schwartz-function that
depends on $|k|$ only. It follows that
\begin{eqnarray}\label{eq:IR1}
|G_x(k,\lambda)-G_0(k,\lambda)| &\leq &  \alpha|k|^{1/2}|x||\kappa(k)|\\
|k|\left|\frac{\partial}{\partial|k|} G_x(k,\lambda)\right| &\leq &
\alpha\expect{x}|k|^{-1/2}f(k)\label{eq:IR2}
\end{eqnarray}
with some Schwartz-function $f$ that depends on $\kappa$
and $\nabla\kappa$. For the definitions of the annihilation operator $a(h)$
and the creation operator $a^{*}(h)$, where $h\in L^2(\R^3;\C^2)$, we refer to \cite{ReedSimon2, Spohn2004}.

The Hamiltonian \eqref{vector-ham} is self-adjoint on
$D(H)=D(-\Delta+H_f)$ and bounded from below \cite{Hiroshima2002}.
We use $E=\inf\sigma(H)$ to denote the lowest point of the
spectrum of $H$ and $\Sigma$ to denote the ionization threshold
\begin{equation}\label{def:Sigma}
   \Sigma = \lim_{R\to\infty}\left(\inf_{\ph\in
   D_R,\, \|\ph\|=1}\sprod{\ph}{H\ph}\right),
\end{equation}
where $D_R:=\{\ph\in D(H)| \chi(|x|\leq R)\ph=0\}$.

Our conjugate operator is the second quantized dilatation
generator
\begin{equation}\label{eq:def-B}
  B = \dGamma(b),\qquad b = \frac{1}{2}(k\cdot y+y\cdot k)
\end{equation}
where $y=i\nabla_k$. By Theorem~\ref{thm:HisC2} of
Section~\ref{sec:local-reg}, the Hamiltonian $H$ is locally of
class $C^2(B)$ on $(-\infty,\Sigma)$. That is, the mapping
\begin{equation}\label{eq:loc-C2}
  s\mapsto e^{-iBs}f(H)e^{iBs}\ph
\end{equation}
is twice continuously differentiable, for every $\ph\in\H$ and
every $f\in C_0^{\infty}(-\infty,\Sigma)$. This makes the conjugate
operator theory in the variant of Sahbani \cite{Sahbani1997}
applicable, and, in particular, it allows one to define the
commutator $[H,iB]$ as a sesquilinear form on
$\cup_{K}E_{K}(H)\H$, the union being taken over all compact
subsets $K$ of $(-\infty,\Sigma)$. We are now prepared to state the main results
of this paper.

\begin{theorem}\label{vector-mourre}
Suppose that $\alpha\ll 1$. Then for any $\sigma \leq \egap/2$
$$
   E_{\Delta}(H-E)[H,iB] E_{\Delta}(H-E) \geq\frac{\sigma}{10} E_{\Delta}(H-E),
$$
where $\Delta=[\sigma/3,2\sigma/3]$.
\end{theorem}

Given Theorem~\ref{vector-mourre}, the remark preceding it, and
the fact that, by Lemma~\ref{lm:E-sigma}, $\Sigma\geq E+\egap/3$
for $\alpha$ small enough, we see that both Hypotheses of
Conjugate Operator Theory (Appendix~\ref{sec:conjugate}) are
satisfied for $\Omega=(E,E+\egap/3)$. This implies that the
consequences, Theorems~\ref{thm:Holder} and
Theorem~\ref{thm:decay-in-time}, of the general theory hold for
the system under investigation, and, thus, it proves
Theorem~\ref{vector-LAP} and Theorem~\ref{vector-t-decay} below.
Alternatively, the first part of Theorem~\ref{vector-LAP} can also be derived from
Theorem~\ref{vector-mourre} using Theorem A.1 of \cite{BFSS}.

\begin{theorem}[\textbf{Limiting absorption principle}]\label{vector-LAP}
Let $\alpha\ll 1$. Then for every
$s>1/2$ and all $\ph,\psi\in \H$ the limits
\begin{equation}\label{eq:B-res}
  \lim_{\eps\to 0} \sprod{\ph}{\expect{B}^{-s}(H-\lambda\pm i\eps)^{-1}
   \expect{B}^{-s}\psi}
\end{equation}
exist uniformly in $\lambda$ in any compact subset of
$(E,E+\egap/3)$. For $s\in (1/2,1)$ the map
\begin{equation}\label{eq:res-Holder}
   \lambda\mapsto \expect{B}^{-s}(H-\lambda\pm i0)^{-1}
   \expect{B}^{-s}
\end{equation}
is (locally) H\"older continuous of degree $s-1/2$ in
$(E,E+\egap/3)$.
\end{theorem}

As a corollary from the finiteness of \eqref{eq:B-res} one can show that
$\expect{B}^{-s}f(H)(H-z)^{-1}f(H)\expect{B}^{-s}$
is bounded on $\C_{\pm}$ for all $f\in C_0^{\infty}(\R)$ with
support in $(E,E+\egap/3)$. This implies $H$-smoothness of
$\expect{B}^{-s}f(H)$ and \emph{local decay}
$$
  \int_{\R}\|\expect{B}^{-s}f(H)e^{-iHt}\ph\|^2dt \leq C\|\ph\|^2.
$$
See \cite{ReedSimon4}, Theorem XIII.25 and its Corollary. From the
H\"older continuity of \eqref{eq:res-Holder} we obtain in addition
a pointwise decay in time (c.f.~Theorem~\ref{thm:decay-in-time}).

\begin{theorem}\label{vector-t-decay}
Let $\alpha\ll 1$ and suppose $s\in (1/2,1)$ and $f\in C_0^{\infty}(\R)$ with $\supp(f)\subset
(E,E+\egap/3)$. Then
\begin{equation*}
  \| \expect{B}^{-s}e^{-iHt}f(H)\expect{B}^{-s}\|
  = O(\frac{1}{t^{s-1/2}}), \qquad (t\to\infty).
\end{equation*}
\end{theorem}

\section{Proof of the Mourre Estimate} \label{sec:qed-mourre}

This section describes the main steps of the proof of Theorem~\ref{vector-mourre}.  Technical
auxiliaries such as the existence of a spectral gap, soft boson
bounds, and the localization of the electron are collected in
Appendix~\ref{appendix:A}.

The proof of Theorem~\ref{vector-mourre} depends, of course, on an
explicit expression for the commutator $[H,iB]$. By
Lemma~\ref{lm:com-H-A} and Proposition~\ref{ad-Bs-H}, we know that
for $f\in C_0^{\infty}(-\infty,\Sigma)$
\begin{align}\nonumber
   f(H)[H,iB]f(H)\quad &=\quad \lim_{s\to 0} f(H)\left[H,\frac{e^{iBs}-1}{s}\right]f(H)\\
   &=\quad f(H)\big(\dGamma(\omega) - \alpha^{3/2}\phi(ibG_{x})\cdot\Pi
-\alpha^{3/2}\Pi\cdot\phi(ibG_{x})\big)
   f(H),\label{to-esti-below}
\end{align}
where the limit is taken in the strong operator topology.
Therefore we may identify $[H,iB]$, as a quadratic form, with
$\dGamma(\omega) - \alpha^{3/2}\phi(ibG_{x})\cdot\Pi
-\alpha^{3/2}\Pi\cdot\phi(ibG_{x})$. One of our main tools
for estimating \eqref{to-esti-below} from below is an infrared
cutoff Hamiltonian $H_{\sigma}$, $\sigma$ as in
Theorem~\ref{vector-mourre}, whose spectral subspaces for energies
close to $\inf\sigma(H_{\sigma})$ are explicitly known (see
Lemma~\ref{lm:qed-main}). A second key tool is the decomposition
of $B$ into two pieces, $B_{\sigma}$ and $B^{\sigma}$. We now
define these operators along with some other auxiliary operators
and Hilbert spaces. As a general rule, we will place the index
$\sigma$ downstairs if only \emph{low-energy} photons are
involved, and upstairs for \emph{high-energy} photons. The fact
that this rule does not cover all cases should not lead to any
confusion.

Let $\chi_0,\chi_{\infty}\in C^{\infty}(\R,[0,1])$, with
$\chi_0=1$ on $(-\infty,1]$, $\chi_{\infty}=1$ on $[2,\infty)$,
and $\chi_0^2+\chi_{\infty}^2\equiv 1$. For a given $\sigma>0$, we
define $\chi_{\sigma}(k) = \chi_{0}(|k|/\sigma)$,
$\chi^{\sigma}(k) = \chi_{\infty}(|k|/\sigma)$,
$\tilde{\chi}^{\sigma}(k)=1-\chi_{\sigma}(k)$, and a Hamiltonian
$H_{\sigma}$ by
\begin{equation}\label{eq:vIR-Ham}
   H_{\sigma}=(p+\alpha^{3/2}A^{\sigma}(\alpha x))^2+V+H_f,
\end{equation}
where $p=-i\nabla_x$ and $A^{\sigma}(\alpha
x)=\phi(\tilde{\chi}^{\sigma}G_{x})$. Let $\F_{\sigma}$ and
$\F^{\sigma}$ denote the symmetric Fock spaces over
$L^2(|k|<\sigma)$ and $L^2(|k|\geq\sigma)$, respectively, and let
$\H^{\sigma}=L^2(\R^3)\otimes \F^{\sigma}$. Then $\H$ is
isomorphic to $\H^{\sigma}\otimes \F_{\sigma}$, and, in the sense
of this isomorphism,
\begin{equation}\label{eq:ham-sum}
    H_{\sigma} = H^{\sigma}\otimes 1 + 1\otimes H_{f,\sigma}.
\end{equation}
Here $H^{\sigma}=H_{\sigma}\upharpoonright\H^{\sigma}$ and
$H_{f,\sigma}=H_f\upharpoonright\F_{\sigma}$.

Next, we split the operator $B$ into two pieces depending on
$\sigma$. To this end we define new cutoff functions
$\eta_{\sigma}=\chi_{2\sigma}$, $\eta^{\sigma}=\chi^{2\sigma}$ and
cut-off dilatation generators
$b_{\sigma}=\eta_{\sigma}b\eta_{\sigma}$,
$b^{\sigma}=\eta^{\sigma}b\eta^{\sigma}$. Since
$\eta_{\sigma}^2+(\eta^{\sigma})^2\equiv 1$ and
$[\eta_{\sigma},[\eta_{\sigma},b]]=0=[\eta^{\sigma},[\eta^{\sigma},b]]$
it follows from the IMS-formula that $b=b_{\sigma}+b^{\sigma}$.
Let $B_{\sigma} = \dGamma(b_{\sigma})$ and $B^{\sigma} =
\dGamma(b^{\sigma})$. Then
$$
    B = B_{\sigma} + B^{\sigma}.
$$

Theorem~\ref{thm:HisC2} implies that $H$ is locally
of class $C^2(B)$, $C^2(B_{\sigma})$ and $C^2(B^{\sigma})$ on
$(-\infty,\Sigma)$. By Lemma~\ref{lm:E-sigma}, $\Sigma-E\geq (2/3)\egap$ for $\alpha$
sufficiently small. It follows that
$(-\infty,\Sigma)\supset (-\infty,E+2/3\egap)$ and hence, arguing
as in \eqref{to-esti-below}, that
\begin{align}\label{eq:vector-C2}
 [H,iB_{\sigma}] &= \dGamma(\eta_{\sigma}^2\omega)
-\alpha^{3/2}\phi(ib_{\sigma}G_{x})\cdot \Pi
-\alpha^{3/2}\Pi\cdot\phi(ib_{\sigma}G_{x})\\
\label{eq:vector-C3}
 [H,iB^{\sigma}] &= \dGamma((\eta^{\sigma})^2\omega)
-\alpha^{3/2}\phi(ib^{\sigma}G_{x})\cdot\Pi
-\alpha^{3/2}\Pi\cdot\phi(ib^{\sigma}G_{x})
\end{align}
in the sense of quadratic forms on the range of $\chi(H\leq
E+\egap/2)$, if $\alpha\ll 1$. Also $H^{\sigma}$ is of class
$C^1(B^{\sigma})$ and
\begin{equation}\label{eq:vector-C4}
  [H^{\sigma},iB^{\sigma}] = \dGamma((\eta^{\sigma})^2\omega)
-\alpha^{3/2}\phi(ib^{\sigma}\tilde{\chi}^{\sigma}G_{x})\cdot\Pi
-\alpha^{3/2}\Pi\cdot\phi(ib^{\sigma}\tilde{\chi}^{\sigma}G_{x})
\end{equation}
on $\chi(H^{\sigma}\leq E+\egap/2)\H^{\sigma}$.

%%%%%%%%%%%%%%%%%%%%%%%%%%%%%%%%%%%%%%%%%%%%%%%%%%%%%%%%%%%%%%%%%%%%%%%%%%%%%%%

As a further piece of preparation we introduce smooth versions of the
energy cutoffs $E_{\Delta}(H-E)$ and
$E_{\Delta}(H_{\sigma}-E_{\sigma})$. We choose $f\in
C_0^{\infty}(\R;[0,1])$ with $f=1$ on $[1/3,2/3]$ and
$\supp(f)\subset [1/4,3/4]$, so that $f_{\Delta}(s):=f(s/\sigma)$
is a smoothed characteristic function of the interval
$\Delta=[\sigma/3,2\sigma/3]$. We define
\begin{equation}\label{smooth-E-proj}
  F_{\Delta} = f_{\Delta}(H-E),\qquad
  F_{\Delta,\sigma} = f_{\Delta}(H_{\sigma}-E_{\sigma}).
\end{equation}

Finally, to simplify notations, we set
$$
     \int dk:= \sum_{\lambda=1,2}\int d^3k
$$
and we suppress the index $\lambda$ in $a_{\lambda}(k)$,
$a_{\lambda}^{*}(k)$, and $G_{x}(k,\lambda)$.

%%%%%%%%%%%%% main Lemma %%%%%%%%%%%%%%%%%%%%%%%%%%%%%%%%

\begin{lemma}\label{lm:qed-main}
If $\alpha\ll 1$ and $\sigma\leq \egap/2$, then
\begin{equation*}
  F_{\Delta,\sigma} = P^{\sigma}\otimes
  f_{\Delta}(H_{f,\sigma}),\qquad \text{w.r.t.}\ \H=\H^{\sigma}\otimes \F_{\sigma},
\end{equation*}
where $P^{\sigma}$ denotes the ground state projection of
$H^{\sigma}$.
\end{lemma}

\begin{proof}
By Theorem~\ref{thm:gap} of Appendix~\ref{appendix:A}, $H^{\sigma}$ has the gap
$(E_{\sigma},E_{\sigma}+\sigma)$ in its spectrum if $\alpha\ll 1$.
Since the support of $f_{\Delta}$ is a subset of $(0,\sigma)$, the assertion
follows.
\end{proof}

\begin{prop}\label{pp3}
Let $[H,iB^{\sigma}]$ be defined by \eqref{eq:vector-C3}. If
$\alpha\ll 1$ and $\sigma\leq \egap/2$, then
$$
   F_{\Delta,\sigma}[H,iB^{\sigma}]F_{\Delta,\sigma} = 0.
$$
\end{prop}

\begin{proof}
From $b^{\sigma}=b^{\sigma}\tilde{\chi}^{\sigma}$,
Equations~\eqref{eq:vector-C3} and \eqref{eq:vector-C4} it follows
that $[H,iB^{\sigma}] = [H^{\sigma},iB^{\sigma}]\otimes 1 $ with
respect to $\H=\H^{\sigma}\otimes \F_{\sigma}$. The statement now
follows from Lemma~\ref{lm:qed-main} and the Virial Theorem
$P^{\sigma}[H^{\sigma},iB^{\sigma}]P^{\sigma}=0$, Proposition~\ref{Virial}.
\end{proof}

\begin{prop}\label{pp3b}
Let $[H,iB_{\sigma}]$ be defined by \eqref{eq:vector-C2}. If
 $\alpha\ll 1$ and $\sigma\leq \egap/2$, then
$$
   F_{\Delta,\sigma}[H,iB_{\sigma}]F_{\Delta,\sigma} \geq \frac{\sigma}{8}F_{\Delta,\sigma}^2.
$$
\end{prop}

\begin{proof}
On the right hand side of \eqref{eq:vector-C2} we move the
creation operators $a^{*}(ib_{\sigma}G_{x})$ to the
left of $\Pi$ and the annihilation operators $a(ib_{\sigma}G_{x})$ to the
right of $\Pi$. Since
$$
   \sum_{j=1}^3\Big([\Pi_j,a^{*}(ib_{\sigma}G_{x,j})] +
   [a(ib_{\sigma}G_{x,j}),\Pi_j]\Big) = 0
$$
we arrive at
\begin{equation}\label{eq:pp3.1}
  [H,iB_{\sigma}] = \dGamma(\eta_{\sigma}^2\omega) - 2\alpha^{3/2}
  a^{*}(ib_{\sigma}G_{x})\cdot\Pi - 2\alpha^{3/2}\Pi\cdot
  a(ib_{\sigma}G_{x}).
\end{equation}
Next, we estimate \eqref{eq:pp3.1} from below using only the fraction
$2\alpha^{3/2}\dGamma(\eta_{\sigma}^2\omega)$ of
$\dGamma(\eta_{\sigma}^2\omega)$ at first. By completing the square we get,
using \eqref{eq:IR2},
\begin{eqnarray}\nonumber
  \lefteqn{\dGamma(\chi_{\sigma}^2\omega) - a^{*}(ib_{\sigma}G_{x})\cdot\Pi -
   \Pi\cdot a(ib_{\sigma}G_{x})}\\ \nonumber
   &=&
   \int\omega\Big[\chi_{\sigma}a^{*}- \omega^{-1}\Pi\cdot\big(ib\chi_{\sigma}
   G_{x}\big)^{*}\Big]
   \Big[\chi_{\sigma}a-\omega^{-1}(ib\chi_{\sigma}G_{x})\cdot \Pi\Big]\,
   dk\\ \nonumber
& &-\sum_{n,m=1}^3\int\Pi_n\frac{(b\chi_{\sigma}G_{x,n})^{*}(b\chi_{\sigma}G_{x,m})}{\omega}\Pi_m\,dk\\
   \label{eq:pp3.2}
&\geq & -\const\ \sigma \sum_{n=1}^{3} \Pi_n \expect{x}^2\Pi_n.
\end{eqnarray}
From \eqref{eq:pp3.1} and \eqref{eq:pp3.2} it follows that
\begin{equation}\label{pp1.3}
  [H,iB_{\sigma}] \geq
  (1-2\alpha^{3/2})\dGamma(\eta_{\sigma}^2\omega)
  -\const\ \alpha^{3/2}\sigma \sum_n \Pi_n \expect{x}^2\Pi_n.
\end{equation}
It remains to estimate
$F_{\Delta,\sigma}\dGamma(\eta_{\sigma}^2\omega)F_{\Delta,\sigma}$
from below and $F_{\Delta,\sigma}\sum_n \Pi_n \expect{x}^2\Pi_n F_{\Delta,\sigma}$
from above. Using that $F_{\Delta,\sigma}=P^{\sigma}\otimes f_{\Delta}(H_{f,\sigma})$, by
Lemma~\ref{lm:qed-main}, and
\begin{equation*}
   \dGamma(\eta_{\sigma}^2\omega) \geq H_{f,\sigma},\qquad
  f_{\Delta}(H_{f,\sigma}) H_{f,\sigma} f_{\Delta}(H_{f,\sigma}) \geq \frac{\sigma}{4}f_{\Delta}^2(H_{f,\sigma}),
\end{equation*}
we obtain
\begin{equation}\label{pp1.5}
   F_{\Delta,\sigma}\dGamma(\eta_{\sigma}^2\omega)F_{\Delta,\sigma} \geq \frac{\sigma}{4} F_{\Delta,\sigma}^2.
\end{equation}
Furthermore, by Lemma~\ref{qed-x-loc} and Lemma~\ref{lm:basic},
\begin{equation}\label{pp1.6}
    \sup_{\sigma>0}\|\expect{x}\Pi E_{[0,\egap/2]}(H_{\sigma}-E_{\sigma})\| <\infty.
\end{equation}
Since $E_{[0,\egap/2]}(H_{\sigma}-E_{\sigma})F_{\Delta,\sigma}=F_{\Delta,\sigma}$
the proposition follows from \eqref{pp1.3}, \eqref{pp1.5}, and
\eqref{pp1.6}.
\end{proof}

%%%%%%%%%%%%%%%%%%%%%%%%%%%  prop qed-2 %%%%%%%%%%%%%%%%%%%%%%%%%%%%%%%%%%%%%%%%%%%%%%%%%%%%%%

\begin{prop}\label{pp4}
Let $F_{\Delta}, F_{\Delta,\sigma}$ be given by \eqref{smooth-E-proj}. 
There exists a constant $C$ such that for $\alpha\ll 1$ and $\sigma\leq \egap/2$,
$$
\big\|F_{\Delta}-F_{\Delta,\sigma}\big\| \leq C\alpha^{3/2}\sigma.
$$
\end{prop}

\begin{proof}
We begin with a Pauli-Fierz transformation $U_{\sigma}$ effecting only the
photons with $|k|\leq \sigma$. Let
\begin{equation*}
  U_{\sigma} = \exp(i\alpha^{3/2}x\cdot A_{\sigma}(0)),\qquad
  A_{\sigma}(\alpha x) := \phi(\chi_{\sigma}G_x).
\end{equation*}
Then
\begin{eqnarray*}
   H_{(\sigma)} &:=& U_{\sigma}H U^{*}_{\sigma}\\
   &=&  \Big(p+\alpha^{3/2}A^{(\sigma)}(\alpha x)\Big)^2 + V + H_f + \alpha^{3/2}x\cdot E_{\sigma}(0)
   +\frac{2}{3}\alpha^3 x^2\|\chi_{\sigma}\kappa\|^2,
\end{eqnarray*}
where $A^{(\sigma)}(\alpha x) := A(\alpha x)- A_{\sigma}(0)$ and $E_{\sigma}(0) := -i[H_f,A_{\sigma}(0)]$.
We compute, dropping the argument $\alpha x$ temporarily,
\begin{equation}\label{pp4.0}
\begin{split}
  H_{(\sigma)}-H_{\sigma} =&\ 2\alpha^{3/2}p\cdot (A^{(\sigma)}-A^{\sigma}) \\
  &+\alpha^3(A^{(\sigma)}+ A^{\sigma})\cdot (A^{(\sigma)} - A^{\sigma})\\
  &+\alpha^{3/2}x\cdot E_{\sigma}(0) + \frac{2}{3}\alpha^{3} x^2\|\chi_{\sigma}\kappa\|^2,
\end{split}
\end{equation}
where \((A^{(\sigma)})^2 - (A^{\sigma})^2=(A^{(\sigma)}
+ A^{\sigma})\cdot (A^{(\sigma)} - A^{\sigma})\) was used. Note that
$A^{(\sigma)}\cdot A^{\sigma}=A^{\sigma}\cdot A^{(\sigma)}$.
For later reference we note that
\begin{gather}\label{pp4.1a}
  A^{(\sigma)}(\alpha x) - A^{\sigma}(\alpha x) = A_{\sigma}(\alpha x)
  - A_{\sigma}(0)=  \phi(\chi_{\sigma}(G_{x}-G_0))\\
  x\cdot E_{\sigma}(0) = \phi(i\omega \chi_{\sigma}G_0\cdot x). \label{pp4.1b}
\end{gather}

\noindent\underline{Step 1}. Uniformly in $\sigma\leq \egap/2$,
\begin{equation}\label{pp4.4}
\|(U_{\sigma}^{*}-1)F_{\Delta,\sigma}\| =
O(\alpha^{3/2}\sigma),\qquad(\alpha\to 0).
\end{equation}

\noindent\emph{Proof of Step 1}. By the spectral theorem
\begin{eqnarray*}
\|(U^{*}_{\sigma}-1)F_{\Delta,\sigma}\| &\leq & \|\alpha^{3/2}x\cdot
A_{\sigma}(0)F_{\Delta,\sigma}\|\\
& = & \alpha^{3/2} \|x\cdot\phi(\chi_{\sigma} G_0)F_{\Delta,\sigma}\|\\
& \leq & 2\alpha^{3/2} \|x\cdot a(\chi_{\sigma} G_0)F_{\Delta,\sigma}\| +
\alpha^{3/2}\|\chi_{\sigma}G_0\|\cdot \|x F_{\Delta,\sigma}\|.
\end{eqnarray*}
The second term is of order $\alpha^{3/2}\sigma$ as $\sigma\to 0$, because, by
assumption on $G_0$, $\|\chi_{\sigma}G_0\|=O(\sigma)$, and because
\(\sup_{0<\sigma\leq\egap/2}\|x F_{\Delta,\sigma}\|<\infty\) by
Lemma~\ref{qed-x-loc}. The first term is of order
$\alpha^{3/2}\sigma$ as well, by Lemma~\ref{lm:overlap} and Lemma~\ref{qed-x-loc}.

\noindent\underline{Step 2}. Let
$F_{\Delta,(\sigma)}:=f_{\Delta}(H_{(\sigma)}-E)=U_{\sigma}F_{\Delta}U_{\sigma}^{*}$.
Then, uniformly in $\sigma\leq \egap/2$,
\begin{equation}\label{pp4.3}
  \|F_{\Delta,(\sigma)}-F_{\Delta,\sigma}\| = O(\alpha^{3/2}\sigma),\qquad(\alpha\to 0).
\end{equation}

Step 1 and Step 2 complete the proof of the proposition, because 
\begin{align*}
  F_{\Delta} - F_{\Delta,\sigma} &= U_{\sigma}^{*}F_{\Delta,(\sigma)}U_{\sigma} - F_{\Delta,\sigma}\\
&= (U^{*}_{\sigma}-1)F_{\Delta,\sigma} +
U^{*}_{\sigma}F_{\Delta,\sigma}(U_{\sigma}-1)
 + U^{*}_{\sigma}\Big(F_{\Delta,(\sigma)} -
 F_{\Delta,\sigma}\Big)U_{\sigma}.
\end{align*}

\noindent\emph{Proof of Step 2}. Let $j\in C_0^{\infty}([0,1],\R)$ with $j=1$ on $[1/4,3/4]$ and
$\supp(j)\subset[1/5,4/5]$. Let $j_{\Delta}(s)=j(s/\sigma)$, so
that $f_{\Delta}j_{\Delta}=f_{\Delta}$, and let
$J_{\Delta}=j_{\Delta}(H-E)$ and
$J_{\Delta,\sigma}=j_{\Delta}(H_{\sigma}-E_{\sigma})$. We will show that
\begin{align}\label{pp4.5}
  \|F_{\Delta,(\sigma)}-F_{\Delta,\sigma}\| &= O(\alpha^{3/2}\sigma^{1/2}),\\
  \label{pp4.6}
   \|(F_{\Delta,(\sigma)}-F_{\Delta,\sigma})J_{\Delta,\sigma}\| &=
   O(\alpha^{3/2}\sigma),
\end{align}
and it will be clear from our proofs that \eqref{pp4.5} and \eqref{pp4.6} hold likewise with $F$ and
$J$ interchanged. These estimates prove the proposition, because
\begin{align*}
F_{\Delta,(\sigma)}-F_{\Delta,\sigma} =&
F_{\Delta,(\sigma)}J_{\Delta,(\sigma)} - F_{\Delta,\sigma}J_{\Delta,\sigma}\\
=& F_{\Delta,\sigma}(J_{\Delta,(\sigma)}-J_{\Delta,\sigma})
+(F_{\Delta,(\sigma)}-F_{\Delta,\sigma})J_{\Delta,\sigma}\\
&+ (F_{\Delta,(\sigma)}-F_{\Delta,\sigma})(J_{\Delta,(\sigma)}-J_{\Delta,\sigma}).
\end{align*}

To prove \eqref{pp4.5} and \eqref{pp4.6} we use the functional
calculus based on the representation
\begin{equation}\label{HS-calculus}
    f(s)=\int\rd \tilde{f}(z)\frac{1}{z-s},\qquad \rd \tilde{f}(z):=
    -\frac{1}{\pi}\frac{\partial\tilde{f}}{\partial\bar{z}}(z)dxdy,
\end{equation}
for an almost analytic extension $\tilde{f}$ of $f$ that satisfies
$|\partial_{\bar{z}}\tilde{f}(x+iy)|\leq \const\ y^2$ \cite{HelSjo1989, dav}.

We begin with the proof of \eqref{pp4.6}. From \eqref{smooth-E-proj} and
\eqref{HS-calculus} we obtain
\begin{multline}\label{pp4.7}
  (F_{\Delta,(\sigma)}-F_{\Delta,\sigma})J_{\Delta,\sigma} \\= \sigma^{-1}\int \rd\tilde{f}(z)\frac{1}{z-(H_{(\sigma)}-E)/\sigma}
  \Big(H_{(\sigma)}-H_{\sigma}-E+E_{\sigma}\Big)J_{\Delta,\sigma}\frac{1}{z-(H_{\sigma}-E_{\sigma})/\sigma}.
\end{multline}
Since, by Lemma~\ref{lm5}, $|E-E_{\sigma}|=O(\alpha^{3/2}\sigma^2)$, it remains
to estimate the contributions of the various terms due to
$H_{(\sigma)}-H_{\sigma}$ as given by \eqref{pp4.0}.
To begin with, we note that
\begin{eqnarray}\label{pp4.8}
   \|(A^{(\sigma)}-A^{\sigma})J_{\Delta,\sigma}\| &=& O(\alpha\sigma^2)\\
   \|x\cdot E_{\sigma}(0) J_{\Delta,\sigma}\| &=& O(\sigma^2).\label{pp4.9}
\end{eqnarray}
This follows from \eqref{pp4.1a}, \eqref{pp4.1b}, \eqref{eq:IR1},
and Lemma~\ref{lm:overlap}, as far as the
annihilation operators in \eqref{pp4.8} and \eqref{pp4.9} are
concerned. For the term due to the creation operator in \eqref{pp4.8}
we use
\[
  \|a^*(\chi_{\sigma}(G_x-G_0))J_{\Delta,\sigma}\|
  \leq \|a(\chi_{\sigma}(G_x-G_0))J_{\Delta,\sigma}\|
  + \big\|\|\chi_{\sigma}(G_x-G_0)\|\, J_{\Delta,\sigma}\big\|
\]
and $\|\chi_{\sigma}(G_x-G_0)\|=O(|x|\alpha\sigma^2)$, as well as
$\sup_{\sigma>0}\| |x|J_{\Delta,\sigma}\|<\infty$.
The operators $p$ and $A^{(\sigma)}+ A^{\sigma}$ stemming from the first and
second terms of \eqref{pp4.0} are combined with the first resolvent of
\eqref{pp4.7}: using $U_{\sigma}^{*}pU_{\sigma}=p+\alpha^{3/2}A_{\sigma}(0)$
and Lemma~\ref{lm:basic} we obtain
\begin{eqnarray*}
\|(z-(H_{(\sigma)}-E)/\sigma)^{-1}p\| &=&
\|(z-(H-E)/\sigma)^{-1}(p+\alpha^{3/2}A_{\sigma}(0))\|\\
&\leq & \const \frac{\sqrt{1+|z|}}{|y|}
\end{eqnarray*}
which is integrable with respect to $\rd\tilde{f}(z)$. This proves that the
first, second and third terms of \eqref{pp4.0} give contributions to
\eqref{pp4.7} of order $\alpha^{5/2}\sigma$, $\alpha^4\sigma$, and
$\alpha^{3/2}\sigma$, respectively. Since
$\|\chi_{\sigma}\kappa\|^2=O(\sigma^3)$, \eqref{pp4.6} follows.

The proof of \eqref{pp4.5} is somewhat involved due to factors of $x$. We
begin with
\begin{align*}
   F_{\Delta,(\sigma)} - F_{\Delta,\sigma} &= F_{\Delta,(\sigma)}J_{\Delta,(\sigma)} - F_{\Delta,\sigma}J_{\Delta,\sigma}\\
   &= (F_{\Delta,(\sigma)} - F_{\Delta,\sigma})J_{\Delta,\sigma} +  F_{\Delta,(\sigma)}(J_{\Delta,(\sigma)}-J_{\Delta,\sigma})
\end{align*}
The first term is of order $\alpha^{3/2}\sigma$ by \eqref{pp4.6}. The
second one can be written as
\begin{equation}\label{pp4.10}
  \sigma^{-1}\int d\tilde{f}(z) R_{(\sigma)}(z)F_{\Delta,(\sigma)}
  \Big(H_{(\sigma)}-H_{\sigma}-E+E_{\sigma}\Big)R_{\sigma}(z),
\end{equation}
with obvious notations for the resolvents. We recall that, by Lemma~\ref{lm5},
$|E-E_{\sigma}|=O(\alpha^{3/2}\sigma^2)$. As in the proof of
\eqref{pp4.6} we need to estimate the contributions due to the four terms of
$H_{(\sigma)}-H_{\sigma}$ given by \eqref{pp4.0}. We do this exemplarily for the
second one and begin with the estimate
\begin{eqnarray}\nonumber
  \lefteqn{\|F_{\Delta,(\sigma)}
 (A^{(\sigma)}+A^{\sigma})\cdot(A^{(\sigma)}-A^{\sigma})R_{\sigma}(z)\|}\\
 & \leq & \|F_{\Delta,(\sigma)}\expect{x}(A^{(\sigma)}+A^{\sigma})\| \|\expect{x}^{-1}(A^{(\sigma)}-A^{\sigma})(H_f+1)^{-1/2}\|
\|(H_f+1)^{1/2}R_{\sigma}(z)\| \label{pp4.11}
\end{eqnarray}
For the second factor of \eqref{pp4.11} we use
\begin{eqnarray*}
  \|\expect{x}^{-1}(A^{(\sigma)}-A^{\sigma})(H_f+1)^{-1/2}\| &=&
  \|\expect{x}^{-1}\phi(\chi_{\sigma}(G_{x}-G_0))(H_f+1)^{-1/2}\|\\
  &\leq & 2\sup_{x}\expect{x}^{-1}\|\chi_{\sigma}(G_x-G_0)\|_{\omega}\\
 &=& O(\alpha\sigma^{3/2}),
\end{eqnarray*}
which is of the desired order. In the first factor of \eqref{pp4.11} we use that
$U_{\sigma}$ commutes with $\expect{x}$, $A^{(\sigma)}$,
and $A^{\sigma}$, as well as Lemma~\ref{lm:a-Hf}, Lemma~\ref{lm:basic} and
Lemma~\ref{qed-x-loc}. We obtain the bound
\begin{eqnarray*}
  \|F_{\Delta,(\sigma)}\expect{x}(A^{(\sigma)}+A^{\sigma})\| & = &
  \|F_{\Delta}\expect{x}(A^{(\sigma)}+A^{\sigma})\|\\
  &\leq & \|F_{\Delta}\expect{x}(H_f+1)^{1/2}\|
  \|(H_f+1)^{-1/2}(A^{(\sigma)}+A^{\sigma})\|\\
  &\leq & \const\ \|F_{\Delta}(\expect{x}^2+H_f+1)\| <\infty.
\end{eqnarray*}
Finally, for the last factor of \eqref{pp4.11}, Lemma~\ref{lm:basic} implies
the bound
$$
   \|(H_f+1)^{1/2}R_{\sigma}(z)\| \leq \const \frac{\sqrt{1+|z|}}{|y|},
$$
which is integrable with respect to $\rd \tilde{f}(z)$. In a
similar way the contributions of the other terms of \eqref{pp4.0}
are estimated. It follows that \eqref{pp4.10} is of order
$O(\alpha^{3/2}\sigma^{1/2})$ which proves \eqref{pp4.5}. This
completes the proof of Proposition~\ref{pp4}.
\end{proof}

\begin{proof}[{\bf Proof of Theorem~\ref{vector-mourre}}]
Since $(\eta^{\sigma})^2+\eta_{\sigma}^2=1$ and $b_{\sigma}+b^{\sigma}=b$, it follows from \eqref{eq:vector-C2}
and \eqref{eq:vector-C3} that
$C:=\dGamma(\omega) - \alpha^{3/2}\phi(ibG_{x})\cdot\Pi
-\alpha^{3/2}\Pi\cdot\phi(ibG_{x})=[H,iB_{\sigma}]+[H,iB^{\sigma}]$. Thus Propositions~\ref{pp3} and \ref{pp3b} imply that
\[
   F_{\Delta,\sigma} C F_{\Delta,\sigma} \geq \frac{\sigma}{8}F_{\Delta,\sigma}^2.
\]
We next replace $F_{\Delta,\sigma}$ by $F_{\Delta}$, using
Proposition~\ref{pp4} and noticing that $CF_{\Delta,\sigma}$ and $F_{\Delta}C$ are bounded,
uniformly in $\sigma$. Since, by \eqref{to-esti-below},
$C=[H,iB]$ on the range of $F_{\Delta}$ we arrive
at
\[
   F_{\Delta}[H,iB] F_{\Delta} \geq \frac{\sigma}{8}F_{\Delta}^2 +
   O(\alpha^{3/2}\sigma).
\]
After multiplying this operator inequality from both sides with
$E_{\Delta}(H-E)$, the theorem follows.
\end{proof}

%%%%%%%%%%%%%%%%%%%%%%%%%%%%%%%%%%%%%%%%%%%%%%%%%%%%%%%%%%%%%%%%%%

\section{Local regularity of $H$ with respect to $B$}
\label{sec:local-reg}

The purpose of this section is to prove that $H$ is locally of
class $C^2(B)$ in $(-\infty,\Sigma)$, where $\Sigma$ is the
ionization threshold of $H$, and $B$ is any of the three operators
$\dGamma(b), \dGamma(b_{\sigma}), \dGamma(b^{\sigma})$ defined in
Section~\ref{sec:vector}. Some background on the concept of local
regularity of a Hamiltonian with respect to a conjugate operator
and basic criteria for this property to hold are collected in
Appendix~\ref{sec:conjugate}. To prove a result that covers the
three aforementioned operators we consider a class of operators
$B$ that contains all of them and is defined as follows.

Let $k\mapsto v(k)$ be a $C^{\infty}$-vector field on $\R^3$ of the form
$v(k)=h(|k|)k$ where $h\in C^{\infty}(\R)$ such that $s^n\partial^n
h(s)$ is bounded for $n\in \{0,1,2\}$. It follows
\begin{equation}
  \label{beta-k}
  |v(k)| \leq \beta |k|,\qquad \text{for all}\ k\in \R^3,
\end{equation}
for some $\beta>0$, and that partial derivatives of $v$ times a
Schwartz-function, such as $\kappa$, are bounded. We remark that the
assumption that $v$ is parallel to $k$ is not needed if a representation of $H$ free of
polarization vectors is chosen.

Let $\phi_s:\R^3\to \R^3$ be the
flow generated by $v$, that is,
\begin{equation}
  \label{flow}
  \frac{d}{ds}\phi_s(k) = v(\phi_s(k)),\qquad \phi_0(k)=k.
\end{equation}
Then $\phi_s(k)$ is of class $C^{\infty}$ with respect to $s$ and $k$, and by
Gronwall's lemma and \eqref{beta-k}
\begin{equation}
  \label{gronwall}
  e^{-\beta|s|}|k| \leq |\phi_s(k)| \leq e^{\beta|s|} |k|,\qquad \text{for}\ s\in \R.
\end{equation}
Induced by the flow $\phi_s$ on $\R^3$ there is a one-parameter group of
unitary transformations on $L^2(\R^3)$ defined by
\begin{equation}
  \label{flow-rep}
  f_{s}(k) = f(\phi_{s}(k))\sqrt{\det D\phi_{s}(k)}.
\end{equation}
Since these transformations leave $C_0^{\infty}(\R^3)$ invariant,
their generator $b$ is essentially self-adjoint on this space.
From $bf=id/ds\, f_s|_{s=0}$ we obtain
\begin{equation}
  \label{gerarator-b}
  b = \frac{1}{2}(v\cdot y + y\cdot v)
\end{equation}
where $y=i\nabla_{k}$. Let $B=\dGamma(b)$. The main result of this
section is:

%%%%%%%%%%  H is locally of class C2 %%%%%%%%%%%%%%%%%%%%%%%

\begin{theorem}\label{thm:HisC2}
Let $H$ be the Hamiltonian defined by \eqref{vector-ham} and let $\Sigma$ be
its ionization threshold given by \eqref{def:Sigma}. Under the assumptions above on the vector-field $v$,
the operator $H$ is
locally of class $C^2(B)$ in $\Omega=(-\infty,\Sigma)$ for all values of $\alpha$.
\end{theorem}

The proof, of course, depends on the explicit knowledge of the unitary group
generated by $B$, and in particular on the formulas
\begin{gather}\label{Hf-flow}
  e^{-iBs} H_f e^{iBs} = \dGamma(e^{-ibs}\omega e^{ibs}) = \dGamma(\omega\circ
  \phi_{s})\\ \label{A-flow}
  e^{-iBs} A(x) e^{iBs} = \phi(e^{-ibs}G_{x}) = \phi(G_{x,s})
\end{gather}
with $G_{x,s}$ given by \eqref{flow-rep}. Another essential ingredient is
that, by \cite{Griesemer2004}, Theorem~1,
\begin{equation}\label{gri-x-loc}
   \|\expect{x}^2 f(H)\| <\infty
\end{equation}
for every $f\in C_0^{\infty}(\Omega)$. We begin with four
auxiliary results, Propositions~\ref{domains}, \ref{ad-Bs-H},
\ref{ad-Bs-F}, and \ref{ad2-Bs-H}.

\begin{prop}\label{domains}
\begin{itemize}
\item[(a)] For all $s\in\R$, $e^{iBs}D(H_f)\subset D(H_f)$ and
$$
   \|H_f e^{iBs}(H_f+1)^{-1}\| \leq e^{\beta|s|}
$$
\item[(b)] For all $s\in\R$, $e^{iBs}D(H)\subset D(H)$ and
 $$
   \|H e^{iBs}(H+i)^{-1}\| \leq \const\ e^{\beta|s|}
$$
\end{itemize}
\end{prop}

\begin{proof}
From $e^{-iBs}H_f e^{iBs} = \dGamma(e^{-ibs}\omega) = \dGamma(\omega\circ
\phi_s)$ and \eqref{gronwall} it follows that
$$
  \|H_f e^{iBs}\ph\| = \|\dGamma(\omega\circ\phi_s)\ph\| \leq e^{\beta|s|}\|H_f\ph\|
$$
for all $\ph\in \F_0(C_0^{\infty})$, which is a core of $H_f$.
This proves, first, that $e^{iBs}D(H_f)\subset D(H_f)$, and next,
that the estimate above extends to $D(H_f)$, proving (a).

The Hamiltonian $H$ is self-adjoint on the domain of
$H^{(0)}=-\Delta+H_f$. Therefore the operators $H^{(0)}(H+i)^{-1}$
and $H(H^{(0)}+i)^{-1}$ are bounded and it suffices to prove (b)
for $H^{(0)}$ in place of $H$. The subspace $D(\Delta)\otimes
D(H_f)$ is a core of $H^{(0)}$. By (a) it is invariant w.r.~to
$e^{iBs}$ and
\begin{equation*}
  \|H^{(0)}e^{iBs}\ph\| \leq  \|\Delta\ph\| + \|H_f\ph\| e^{\beta|s|} \leq \sqrt{2} e^{\beta|s|} \|H^{(0)}\ph\|
\end{equation*}
As in the proof of (a), it now follows that $e^{iBs}D(H^{(0)})\subset
D(H^{(0)})$ and then the estimate above extends to
$D(H^{(0)})$.
\end{proof}

Let $B_s:=(e^{iBs}-1)/is$. Then, by Proposition~\ref{domains},
$[B_s,H]$ is well defined, as a linear operator on $D(H)$. The
main ingredients for the proof of Theorem~\ref{thm:HisC2} are
Propositions~\ref{ad-Bs-H} and \ref{ad2-Bs-H} below.

\begin{prop}\label{ad-Bs-H}
\begin{itemize}
\item[(a)] For all $\ph\in D(H)$
$$
   i\lim_{s\to 0}\expect{x}^{-1}[H,B_s]\ph
   = \expect{x}^{-1}\Big(\dGamma(\nabla\omega\cdot v)
   -\alpha^{3/2}\phi(ibG_{x})\cdot\Pi
   - \Pi\cdot \phi(ibG_{x})\alpha^{3/2}\Big)\ph.
$$
\item[(b)]
\[
    \sup_{0<|s|\leq 1} \|\expect{x}^{-1}[B_s,H](H+i)^{-1}\| <\infty.
\]
\end{itemize}
\end{prop}

\begin{proof}
Part (b) follows from (a) and the uniform boundedness principle.
Part (a) is equivalent to the limit
$$
   i\lim_{s\to 0}\expect{x}^{-1}\frac{1}{s}\left(e^{-iBs}He^{iBs}-H\right)\ph
$$
being equal to the expression on the right hand side of (a). By
\eqref{Hf-flow}, for all $\ph\in D(H_f)$
$$
  \lim_{s\to 0}\frac{1}{s}\left(e^{-iBs}H_f e^{iBs}-H_f\right)\ph
  = \lim_{s\to 0}\frac{1}{s}\dGamma(\omega\circ \phi_{s}-\omega)\ph =
  \dGamma(\nabla\omega\cdot v)\ph,
$$
where the last step is easily established using Lebesgue's
dominated convergence Theorem. The necessary dominants are
obtained from \(|s^{-1}(\omega\circ\phi_{s}-\omega)| \leq
|s|^{-1}(e^{\beta|s|}-1)\omega\), by \eqref{gronwall}, and from
the assumption $\ph\in D(\dGamma(\omega))$.

It remains to consider the contribution due to $H_{\rm int}:=
2\alpha^{3/2}A(\alpha x)\cdot p + \alpha^3A(\alpha x)^2$. Let $\Delta G_{x,s}:=G_{x,s}-G_{x}$. By
\eqref{A-flow},
\begin{eqnarray}\nonumber
  \lefteqn{e^{-iBs} H_{\rm int} e^{iBs} - H_{\rm int}}\\ &=&
  2\alpha^{3/2}\phi(\Delta G_{x,s})\cdot p + \alpha^3\phi(\Delta G_{x,s})\cdot\phi(G_{x})
  + \alpha^3\phi(G_{x,s})\cdot\phi(\Delta G_{x,s}),\label{eq:BsH1}
\end{eqnarray}
a sum of three operators, each of which contains $\Delta G_{x,s}$.
By Lemma~\ref{lm:dilat-L-omega} at the end of this section, for
each $x\in \R^3$
\begin{equation}\label{eq:BsH2}
  \frac{1}{s}\Delta G_{x,s} =
  \frac{1}{s}\left(G_{x,s}-G_{x}\right)\to -ib G_x,\qquad (s\to 0)
\end{equation}
in the norm $\|\cdot\|_{\omega}$ of $L_{\omega}(\R^3)$ (see
Appendix~\ref{appendix:A}), and
\begin{equation}\label{eq:BsH3}
   \sup_{x\in\R^3} \expect{x}^{-1}\|bG_{x}\|_{\omega} < \infty
\end{equation}
by the assumptions on $G_x$. Since the operators $p(H_f+1)^{1/2}(H+i)^{-1}$ and $H_f(H+i)^{-1}$
are bounded by Lemma~\ref{lm:basic} and since, by Lemma~\ref{lm:a-Hf}, $\|\phi(f)(H_f+1)^{-1/2}\|\leq
\|f\|_{\omega}$ and $\|\phi(f)\phi(g)(H_f+1)^{-1}\|\leq
8\|f\|_{\omega}\|g\|_{\omega}$ for all $f,g\in L^2(\R^3)$, it
follows from \eqref{eq:BsH1}, \eqref{eq:BsH2}, and \eqref{eq:BsH3}
that
\begin{eqnarray*}\label{eq:BsH4}
  \lefteqn{\lim_{s\to 0}\expect{x}^{-1}
  \frac{1}{s}\left(e^{-iBs} H_{\rm int} e^{iBs} - H_{\rm
  int}\right)\ph}\\
   &=& \Big(2\alpha^{3/2}\phi(-ibG_x)\cdot p +\alpha^3\phi(-ibG_x)\cdot \phi(G_x)+
  \alpha^3\phi(G_x)\cdot\phi(-ibG_x)\Big)\ph\\
   &=& -\alpha^{3/2}\Big(\phi(ibG_x)\cdot \Pi +\Pi\cdot \phi(ibG_x)\Big)\ph
\end{eqnarray*}
for all $\ph\in D(H)$.
\end{proof}

\begin{prop}\label{ad-Bs-F}
For all $f\in C_0^{\infty}(\Omega)$,
\[
    \sup_{0<|s|\leq 1} \|[B_s, f(H)]\| <\infty.
\]
\end{prop}

\noindent\emph{Remark.} By Proposition~\ref{C1-tool} this Proposition implies that $f(H)$
is of class $C^1(B)$ for all $f\in C_0^{\infty}(\Omega)$.

\begin{proof}
Let $F=f(H)$ and let $\ad{B_s}{F}=[B_s,F]$. If $g\in
C_0^{\infty}(\Omega)$ is such that $g\equiv 1$ on $\supp(f)$ and
$G=g(H)$, then $F=GF$ and hence
$$
   \ad{B_s}{F} = G\ad{B_s}{F} + \ad{B_s}{G}F.
$$
The norm of $\ad{B_s}{G}F$ is equal to the norm of its adjoint
which is $-F^*\ad{B_{-s}}{G^*}$ where $F^*=\bar{f}(H)$ and
$G^*=\bar{g}(H)$. It therefore suffices to prove that
\begin{equation}
   \sup_{0<|s|\leq 1}\|G\ad{B_s}{F}\|<\infty
\end{equation}
for all $f,g\in C_0^{\infty}(\Omega)$. To this end we use the representation
$f(H) = \int\rd\tilde{f}(z) R(z)$ where $R(z)=(z-H)^{-1}$ and $\tilde{f}$ is an almost analytic extension of $f$ with
$|\partial_{\bar{z}}\tilde{f}(x+iy)|\leq \const|y|^2$, c.f. \eqref{HS-calculus}. It follows that
$$
   G\ad{B_s}{F} = \int\rd\tilde{f}(z) R(z) G[B_s,H]R(z),
$$
which is well-defined by Proposition~\ref{domains}, part (b). Upon
writing $[B_s,H]=\expect{x}\expect{x}^{-1}[B_s,H]R(i)(i-H)$ we can
estimate the norm of the resulting expression for $G\ad{B_s}{F}$
with $0<|s|\leq 1$, by
\begin{equation*}
   \|G\ad{B_s}{F}\| \leq \sup_{0<|s|\leq 1}\|\expect{x}^{-1}[B_s,H]R(i)\|
   \|g(H)\expect{x}\|
   \int|\rd\tilde{f}(z)| \|R(z)\| \|(i-H)R(z)\|.
\end{equation*}
Since
\begin{equation}\label{Hres}
  \|(i-H)R(z)\| \leq \const\left(1+\frac{1}{|\Ima(z)|}\right),
\end{equation}
the integral is finite by choice of $\tilde{f}$. The factors in
front of the integral are finite by Proposition~\ref{ad-Bs-H} and
by \eqref{gri-x-loc}.
\end{proof}

%%%%%%%%%%%%%%%%%%%%%%%%%%%%%%%%%%%%%%%%%%%%%%%%%%%%%%%%%%%%%%%%%%%%%%%%%%%%%%
\begin{prop}\label{ad2-Bs-H}
$$
    \sup_{0<|s|\leq 1}\|\expect{x}^{-2}[B_s[B_s,H]](H+i)^{-1}\| <
    \infty.
$$
\end{prop}

\begin{proof}
By Definition of $H$,
\begin{eqnarray*}
  [B_s,[B_s,H]] &=& [B_s,[B_s,H_f]] +
  \alpha^{3/2}[B_s,[B_s,p\cdot\phi(G_x)]]\\
  &  &+ \alpha^3 [B_s,[B_s,\phi(G_x)^2]].
\end{eqnarray*}
We estimate the contributions of these terms one by one in
Steps~1-3 below. As a preparation we note that
\begin{eqnarray}\label{eq:ad1-W}
   \mathrm{ad}_{B_s} &=& i e^{iBs}\frac{1}{s}(W(s)-1)\\
   \mathrm{ad}_{B_s}^2 &=& -e^{2iBs} \frac{1}{s^2}(W(s)-1)^2 = -e^{2iBs}
    \frac{1}{s^2}\big(W(2s)-2W(s)+W(0)\big),\label{eq:ad2-W}
\end{eqnarray}
Where $W(s)$ maps an operator $T$ to $e^{-iBs}T e^{iBs}$.
In view of Equations~\eqref{Hf-flow}, \eqref{A-flow}, we will need
that for every twice differentiable function $f:[0,2s]\to\C$
\begin{equation}\label{DC1}
  \frac{1}{s^2}|f(2s)-2f(s)+f(0)| \leq \sup_{|t|\leq
    2|s|}|f''(t)|.
\end{equation}

\noindent\underline{Step 1}.
$$
      \sup_{|s|\leq 1}\|\mathrm{ad}^2_{B_s}(H_f)(H_f+1)^{-1}\|<\infty.
$$
By \eqref{eq:ad2-W} and \eqref{Hf-flow}
\begin{equation}\label{DC1.1}
    \mathrm{ad}^2_{B_s}(H_f)= -e^{2iBs}\frac{1}{s^2}
    \dGamma(\omega\circ\phi_{2s}-2\omega\circ\phi_s+\omega).
\end{equation}
Thus in view of \eqref{DC1} we estimate the second derivative of
$s\mapsto \omega\circ\phi_s(k)=|\phi_s(k)|$. For $k\neq 0$,
\begin{eqnarray*}
\frac{\partial^2}{\partial s^2}|\phi_s(k)| &=& -
\frac{1}{|\phi_s(k)|}\sprod{\phi_s(k)}{v(\phi_s(k))}^2 +
\frac{v(\phi_s(k))}{|\phi_s(k)|}\\
&& + \frac{1}{|\phi_s(k)|}\sum_{i,j}\phi_s(k)_i
v_{i,j}(\phi_s(k))\phi_s(k)_j.
\end{eqnarray*}
By assumption on $v$, $v_{i,j}\in L^{\infty}$ and
$|v(\phi_s(k))|\leq \beta|\phi_s(k)|\leq e^{\beta|s|}|k|$. It
follows that
$$
   \frac{1}{s^2}\big|\big(\omega\circ\phi_{2s}-2\omega\circ\phi_s+\omega\big)(k)\big|
   \leq  \const\ e^{\beta|s|}\omega(k),
$$
which implies
$$
   \left\|\frac{1}{s^2}\dGamma(\omega\circ\phi_{2s}-2\omega\circ\phi_s+
   \omega)(H_f+1)^{-1}\right\| \leq \const\ e^{\beta|s|}.
$$
By \eqref{DC1.1} this establishes Step 1.\\

\noindent\underline{Step 2}.
$$
  \sup_{|s|\leq 1}\sup_{x\in\R^3}\expect{x}^{-2}\|\mathrm{ad}^2_{B_s}(\phi(G_x)\cdot p)(H+i)^{-1}\|<\infty.
$$
Since $p(H_f+1)^{1/2}(H+i)^{-1}$ is bounded, it suffices to show
that
\begin{equation}\label{DC2.0}
  \sup_{|s|\leq 1,\, x}\expect{x}^{-2}\|\mathrm{ad}^2_{B_s}(\phi(G_x))(H_f+1)^{-1/2}\|<\infty.
\end{equation}
By Equation~\eqref{A-flow}
\begin{equation}\label{DC2.1}
  \frac{1}{s^2}(W(s)-1)^2(\phi(G_x)) =
  \frac{1}{s^2}\phi(G_{x,2s}-2G_{x,s}+G_{x}),
\end{equation}
and by \eqref{DC1}
\begin{eqnarray*}
 && \expect{x}^{-2}\frac{1}{s^2}\left\|\phi(G_{x,2s}-2G_{x,s}+G_{x})(H_f+1)^{-1/2}\right\|\\
&\leq
&\expect{x}^{-2}\frac{1}{s^2}\|G_{x,2s}-2G_{x,s}+G_{x}\|_{\omega}
\ \leq\ \expect{x}^{-2} \left\|\frac{\partial^2}{\partial
s^2}G_{x,s}\right\|_{\omega}
\end{eqnarray*}
For $k\neq 0$ the function $s\mapsto G_{x,s}(k)$ is arbitrarily
often differentiable by assumption on $v$ and
\begin{eqnarray}\label{DC2.1b}
   -i\frac{\partial}{\partial s}G_{x,s}(k) &=& (v\cdot \nabla_k
   G_x)_s(k) + \frac{1}{2}(\div(v)G_x)_s(k)\\
   -\frac{\partial^2}{\partial s^2}G_{x,s}(k) &=& \big((v\cdot
   \nabla_k)^2 G_x\big)_s(k) + (\div(v)v\cdot\nabla_{k}G_x)_s\label{DC2.2}\\
   && +\frac{1}{2}\sum_{i,j}\big((v_i\partial_i\partial_j v_j)G_x\big)_s +
   \frac{1}{4} \big(\div(v)^2G_x\big)_s.\label{DC2.3}
\end{eqnarray}
By part (a) of Lemma~\ref{lm:dilat-L-omega} below, it suffices to
estimate the $L^2_{\omega}$-norm of these four contributions with
$s=0$. By our assumptions on $v$, $\div(v)$ and
$v_i\partial_i\partial_j v_j$ are bounded functions. This and the
bound $\|G_x\|\leq \|G_0\|_{\omega}<\infty$ account for the
contributions of \eqref{DC2.3}, and for the factor $\div(v)$ in
front of the second term of \eqref{DC2.2}. It remains to show that
the $L^2_{\omega}$-norms of
$$
    \expect{x}^{-1}(v\cdot\nabla_k)G_x\quad\text{and}\quad \expect{x}^{-2}(v\cdot\nabla_k)^2G_x
$$
are bounded uniformly in $x$. But this is easily seen by applying
$v\cdot\nabla_k$ to each factor of $G_x(k,\lambda) =
\eps_{\lambda}(k)e^{-ik\cdot x}\kappa(k)|k|^{-1/2}$ and using that
$v\cdot\nabla \eps_{\lambda}(k)=0$, $v\cdot\nabla e^{-ik\cdot
x}=-iv\cdot x e^{-ik\cdot x}$ and that $v\cdot\nabla|k|^{-1/2}$ is
again of order $|k|^{-1/2}$ by assumption on $v$.

 \noindent\underline{Step 3}.
$$
  \sup_{|s|\leq 1,\, x}\expect{x}^{-2}\|\mathrm{ad}^2_{B_s}(\phi(G_x)^2)(H_f+1)^{-1}\|<\infty.
$$

By the Leibniz-rule for $\mathrm{ad}_{B_s}$,
\begin{eqnarray}\label{DC3.1}
   \mathrm{ad}^2_{B_s}(\phi(G_x)^2) &=&
   \mathrm{ad}^2_{B_s}(\phi(G_x))\cdot
   \phi(G_x)+\phi(G_x)\cdot\mathrm{ad}^2_{B_s}(\ph(G_x))\\
     && +2\ad{B_s}{\phi(G_x)}\ad{B_s}{\phi(G_x)}.\nonumber
\end{eqnarray}
For the contribution of the first term we have
\begin{eqnarray*}
  \lefteqn{\expect{x}^{-2}\|\mathrm{ad}^2_{B_s}(\phi(G_x))\cdot
  \phi(G_x)(H_f+1)^{-1}\|}\\ && \leq \expect{x}^{-2}\|\mathrm{ad}^2_{B_s}(\phi(G_x))(H_f+1)^{-1/2}\|
  \|\phi(G_x)(H_f+1)^{-1/2}\|
\end{eqnarray*}
which is bounded uniformly in $|s|\leq 1$ and $x\in \R^3$ by \eqref{DC2.0} in the
proof of Step~2. For
the second term of \eqref{DC3.1} we first note that
\begin{eqnarray*}
     \phi(G_x)\mathrm{ad}^2_{B_s}(\phi(G_x)) &=& \phi(G_x)
     e^{2iBs}\frac{1}{s^2}(W(s)-1)^2(\phi(G_x))\\
     &=&  e^{2iBs}\phi(G_{x,s})\frac{1}{s^2}(W(s)-1)^2(\phi(G_x))
\end{eqnarray*}
and hence, by the estimates in Step~2, we obtain a bound similar to the one
for the first term of \eqref{DC3.1} with an additional factor of $e^{2\beta|s|}$ coming from the use of
Lemma~\ref{lm:dilat-L-omega}. Finally, by \eqref{eq:ad1-W} and
\eqref{A-flow}
\begin{equation*}
\ad{B_s}{\phi(G_x)}\ad{B_s}{\phi(G_x)} =
e^{2iBs}\phi\left(\frac{G_{x,2s}-G_{x,s}}{s}\right)\phi\left(\frac{G_{x,s}-G_{x}}{s}\right)
\end{equation*}
which implies that
$$
   \expect{x}^{-2}\|\ad{B_s}{\phi(G_x)}\ad{B_s}{\phi(G_x)}(H_f+1)^{-1}\|
   \leq \sup_{|s|\leq 2,\,x\in\R^3}\big(\expect{x}^{-1}\|\partial_s
   G_{x,s}\|_{\omega}\big)^2.
$$
This is finite by \eqref{DC2.1b} and the assumptions on $v$ and $G_x$.
\end{proof}

\begin{proof}[\textbf{Proof of Theorem~\ref{thm:HisC2}}]
By Proposition~\ref{ad-Bs-F} and \ref{C2-tool} it suffices to show
that
\begin{equation}\label{reg1}
  \sup_{0<s\leq 1}\|\add{B_s}{f(H)}\| <\infty
\end{equation}
for all $f\in C_0^{\infty}(\Omega)$. Let $g\in
C_0^{\infty}(\Omega)$ with $gf=f$ and let $G=g(H)$, $F=f(H)$. Then
$F=GF$ and hence
\begin{equation*}
 \add{B_s}{F} = \add{B_s}{GF}
  = \add{B_s}{G}F + 2\ad{B_s}{G}\ad{B_s}{F}+G\add{B_s}{F}.
\end{equation*}
From Proposition~\ref{ad-Bs-F} we know that $\sup_{0<s\leq
1}\|\ad{B_s}{G}\| <\infty$, and similarly with $F$ in place of
$G$. Moreover
$$
  \Big(\add{B_s}{G}F\Big)^{*} = F^{*}\add{B_{-s}}{G^{*}}.
$$
Thus it suffices to show that for all $g,f\in
C_0^{\infty}(\Omega)$
\begin{equation}\label{reg2}
  \sup_{0<|s|\leq 1}\|G\add{B_s}{F}\| <\infty.
\end{equation}
To this end we use $F=\int\rd\tilde{f}(z)R(z)$ with an almost analytic
extension $\tilde{f}$ of $f$ such that $|\partial_{\bar{z}}\tilde{f}(x+iy)|\leq \const\ |y|^4$.
We obtain
\begin{eqnarray}\label{reg3}
  G\add{B_s}{F} &=& 2\int\rd\tilde{f}(z)R(z)G[B_s,H]R(z)[B_s,H]R(z)\\
                & & +\int\rd\tilde{f}(z)R(z)G[B_s,[B_s,H]]R(z).\label{reg3b}
\end{eqnarray}
Since, by \eqref{gri-x-loc}, $\|G\expect{x}^2\|<\infty$ the norm of the
second term is bounded uniformly in $s\in\{0<|s|\leq 1\}$ by Proposition~\ref{ad2-Bs-H}.
In view of Proposition~\ref{ad-Bs-H} we rewrite \eqref{reg3} (times $1/2$)  as
\begin{eqnarray*}
  \lefteqn{\int\rd\tilde{f}(z)R(z)G\expect{x}[B_s,H]R(z)\expect{x}^{-1}[B_s,H]R(z)} \\
   && -\int\rd\tilde{f}(z)R(z)G\Big[\expect{x},[B_s,H]R(z)\Big]\expect{x}^{-1}[B_s,H]R(z).
\end{eqnarray*}
For the norm of the first integral we get the bound
$$
  \int|\rd\tilde{f}(z)|\|R(z)\|\|G\expect{x}^2\|\|\expect{x}^{-1}[B_s,H]R(i)\|^2
  \|(i-H)R(z)\|^2,
$$
which is bounded uniformly in $s$, by Lemma~\ref{ad-Bs-H}, the
exponential decay on the range of $G=g(H)$ and by construction of
$\tilde{f}$. The norm of the second term is bounded by
\begin{equation}\label{reg4}
  \int|\rd\tilde{f}(z)|\|R(z)\|\,\|g(H)\expect{x}\|\,
  \|\expect{x}^{-1}\big[\expect{x},[B_s,H]R(z)\big]\|\,\|\expect{x}^{-1}[B_s,H]R(z)\|.
\end{equation}
The last factor is bounded by $\|(i-H)R(z)\|$, uniformly in
$s\in(0,1]$, by Proposition~\ref{ad-Bs-H}. For the term in the
third norm we find, using the Jacobi identity and
$[B_s,\expect{x}]=0$, that
\begin{equation}\label{reg5}
  \expect{x}^{-1}\big[\expect{x},[B_s,H]R(z)\big] =
  \expect{x}^{-1}\big[B_s,[\expect{x},H]\big]R(z)+\expect{x}^{-1}[B_s,H]R(z)[\expect{x},H]R(z)
\end{equation}
where
\begin{equation}\label{reg6}
  [\expect{x},H] = 2i \frac{x}{\expect{x}}(p+A) +
  \frac{2}{\expect{x}}+\frac{1}{\expect{x}^3}.
\end{equation}
Since \eqref{reg6} is bounded w.r.to $H$, the norm of the second
term of \eqref{reg5}, by Proposition~\ref{ad-Bs-H}, is bounded by
$\|(i-H)R(z)\|^2$ uniformly in $s$. As for the first term of
\eqref{reg5}, in view of \eqref{reg6}, its norm is estimated like
the norm of $\expect{x}^{-1}[B_s,H]R(z)$ in Lemma~\ref{ad-Bs-H},
which leads to a bound of the form $\const\|(i-H)R(z)\|$. By
\eqref{Hres} and by construction of $\tilde{f}$ it follows that
\eqref{reg4} is bounded uniformly in $|s|\in (0,1]$.
\end{proof}

%%%%%%%%%%%%%%%%%%%%%%%%%%%%%%%%%%%%%%%%%%%%%%%%%%%%%%%%%%%%%%%%%%%%%%%%%%%%%%%%%%%
We conclude this section with a lemma used in the proofs of
Propositions~\ref{ad-Bs-H} and \ref{ad2-Bs-H} above. For the
definition of $L^2_{\omega}(\R^3)$ and its norm see
Appendix~\ref{appendix:A}.

\begin{lemma}\label{lm:dilat-L-omega}
Let $f\mapsto f_s=e^{-ibs}f$ on $L^2_{\omega}(\R^3)$ be defined by
\eqref{beta-k}, \eqref{flow} and \eqref{flow-rep}. Then
\begin{itemize}
\item[(a)] The transformation $f\mapsto f_s$ maps $L^2_{\omega}(\R^3)$
into itself and, for all $s\in\R$,
$$
   \|f_s\|_{\omega} \leq e^{\beta|s|/2}\|f\|_{\omega}.
$$
\item[(b)] The mapping $\R\to L^2_{\omega}(\R^3)$, $s\mapsto f_s$
is continuous.
\item[(c)] For all $f\in L^2_{\omega}(\R^3)$ for which $|k|\mapsto
f(|k|\hat{k})$, $\hat{k}\in \R^3$, is continuously differentiable
on $\R_{+}$ and $\sqrt{\omega}\partial_{|k|}f,
\omega\partial_{|k|}f \in L^2(\R^3)$,
$$
    L^2_{\omega}-\lim_{s\to 0}\frac{1}{s}(f_s-f) = v\cdot \nabla
    f + \frac{1}{2}\mathrm{div}(v)f.
$$
\end{itemize}
\end{lemma}

\noindent\emph{Remark.} Statement (c) shows, in particular, that
$f\in D(b)$ and that $-ibf=v\cdot\nabla f + (1/2)\mathrm{div}(v)f$
for the class of functions $f$ considered there.

\begin{proof}
(a) Making the substitution $q=\phi_s(k)$, $dq=\det D\phi_s(k)dk$
and using \eqref{gronwall} we get
\begin{eqnarray*}
   \|f_s\|^2 &=& \int (|k|^{-1}+1) |f(\phi_s(k))|^2\det
   D\phi_s(k)\, dk\\
   &=& \int (|\phi_{-s}(q)|^{-1}+1) |f(q)|^2\, dq\ \leq\
   e^{\beta|s|}\|f\|^2_{\omega}.
\end{eqnarray*}

(b) For functions $f\in L^2_{\omega}(\R^3)$ that are continuous
and have compact support $\|f_s-f\|_{\omega}\to 0$ follows from
$\lim_{s\to 0}f_s(k)= f(k)$, for all $k\in \R^3$ by an application
of Lebesgue's dominated convergence theorem. From here, (b)
follows by an approximation argument using (a).

(c) By assumption on $f$,
$$
   \tilde{f} := v\cdot \nabla f +\frac{1}{2}\mathrm{div}(v)
   f\in L^2_{\omega}(\R^3).
$$
Using that
$$
    f_s(k)-f(k) = \int_0^s (\tilde{f})_t(k)\, dt,\qquad k\neq 0
$$
and Jensen's inequality we get
\begin{eqnarray*}
  \|s^{-1}(f_s-f)-\tilde{f}\|^2_{\omega} &=& \int dk(|k|^{-1}+1)
  \left|\frac{1}{s}\int_0^s[\tilde{f}_t(k)-\tilde{f}(k)]\, dt \right|^2\\
  &\leq & \int dk
  (|k|^{-1}+1)\frac{1}{s}\int_0^s\left|\tilde{f}_t(k)-\tilde{f}(k)\right|^2dt\\
  &=& \frac{1}{s} \int_0^s\|\tilde{f}_t-\tilde{f} \|^2 dt
\end{eqnarray*}
which vanishes in the limit $s\to 0$ by (b).
\end{proof}

%%%%%%%%%%%%%%%%%%%%%%%%%%%%%%%% Appendix %%%%%%%%%%%%%%%%%%%%%%%%%%%%%%%%%%%%%%%%

\appendix
\section{Operator and Spectral Estimates}
\label{appendix:A}

Let $L^2_{\omega}(\R^3,\C^2)$ denote the linear space of
measurable functions $f:\R^3\to\C^2$ with
$$
  \|f\|^2_{\omega} = \sum_{\lambda=1,2}\int |f(k,\lambda)|^2 (|k|^{-1}+1)d^3k <\infty.
$$
\begin{lemma}\label{lm:a-Hf}
For all $f,g\in L^2_{\omega}(\R^3,\C^2)$
\begin{eqnarray*}
  \|a^{\sharp}(f)(H_f+1)^{-1/2}\| &\leq & \|f\|_{\omega},\\
   \|a^{\sharp}(f)a^{\sharp}(g)(H_f+1)^{-1}\| &\leq & 2\|f\|_{\omega}\|g\|_{\omega},
\end{eqnarray*}
where $a^{\sharp}$ may be a creation or an annihilation operator.
\end{lemma}
\noindent The first estimate of Lemma~\ref{lm:a-Hf} is well known, see e.g., \cite{BFS1}. For a proof of the
second one see \cite{FGS1}.

\begin{lemma}[Operator Estimates]\label{lm:basic}
Let $c_n(\kappa) = \int |\kappa(k)|^2 |k|^{n-3}\, d^3k$ for $n\geq
1$. Then
\begin{alignat*}{2}
(i)&\quad &  A(x)^2 &\leq  8 c_1(\kappa) H_f + 4 c_2(\kappa),\\
(ii)&\quad &  -\frac{8}{3} c_1(\kappa)\alpha^3 p^2 &\leq 2p\cdot A(\alpha x) \alpha^{3/2} + H_f,\\
(iii)&\quad & p^2 &\leq 2\Pi^2 + 2\alpha^3 A(\alpha x)^2.
\end{alignat*}
If $\pm V\leq \eps p^2+b_{\eps}$ for all $\eps>0$, and
if $\eps\in(0,1/2)$ is so small that $16\eps\alpha^3 c_1(\kappa)<1$, then
\begin{alignat*}{2}
(iv)&\quad & \Pi^2 &\leq\frac{1}{1-2\eps}(H+b_{\eps}+8\eps\alpha^2
c_2(\kappa)),\\
(v)&\quad & H_f &\leq
\frac{1}{1-16\eps\alpha^2c_1(\kappa)}(H+b_{\eps}+8\eps\alpha^2 c_2(\kappa)),\\
(vi)&\quad & A(x)^2 &\leq
\frac{8c_1(\kappa)}{1-16\eps\alpha^2c_1(\kappa)}(H+b_{\eps}+8\eps\alpha^2c_2(\kappa))
+4c_2(\kappa).
\end{alignat*}
\end{lemma}

\begin{proof}
Estimate (i) is proved in \cite{GLL}. (ii) is easily derived by
completing the square in creation and annihilation operators, and (iii) follows
from $2\alpha^3p\cdot A(\alpha x)\geq -(1/2)p^2-2\alpha^3 A(\alpha x)^2$.

From the assumption on $V$ and statements (i) and (iii) it follows that
\begin{eqnarray*}
    H &\geq & \Pi^2 - \eps p^2 -b_{\eps} + H_f\\
      &\geq & (1-2\eps) \Pi^2 -2\eps\alpha^3 A(x)^3 + H_f -
      b_{\eps}\\
      &\geq & (1-2\eps) \Pi^2 + (1-16\eps\alpha^3c_{1}(\kappa))H_f
      -8\eps\alpha^3 c_2(\kappa) - b_{\eps},
\end{eqnarray*}
which proves (iv) and (v). Statement (vi) follows from (i) and (v).
\end{proof}

Let $E_{\sigma}=\inf\sigma(H_{\sigma})$ and let
$\Sigma_{\sigma}=\lim_{R\to\infty}\Sigma_{\sigma,R}$
be the ionization threshold for $H_{\sigma}$, that is,
\[
  \Sigma_{\sigma,R} = \inf_{\ph\in D_R,\, \|\ph\|=1}\sprod{\ph}{H_{\sigma}\ph}
\]
where $D_R=\{\ph\in D(H_{\sigma})| \chi(|x|\leq R)\ph=0\}$.

\begin{lemma}[Estimates for $E_{\sigma}$ and $\Sigma_{\sigma}$]\label{lm:E-sigma}
With the above definitions
\begin{enumerate}
\item For all $\alpha\geq 0$, \[  E_{\sigma} \leq e_1 + 4c_{2}(\kappa)\alpha^3. \]
\item If $c_1(\kappa)\alpha^3\leq 1/8$ then
\[ \Sigma_{\sigma,R}\geq e_2-o_R(1) - c_1(\kappa)\alpha^3 C,\quad
(R\to\infty), \]
where $C$ and $o_R(1)$ depend on properties of $\Hpart$ only. In
particular
\[ \Sigma_{\sigma} \geq e_2 - c_1(\kappa)\alpha^3 C \]
uniformly in $\sigma\geq 0$.
\end{enumerate}
\end{lemma}

\begin{proof}
Let $\psi_1$ be a normalized ground state vector of $\Hpart$, so that $\Hpart\psi_1=e_1\psi_1$,
and let $\Omega\in \F$ denote the vacuum. Then
\begin{eqnarray*}
E_{\sigma} & \leq &\sprod{\psi_1\otimes\Omega}{H_{\sigma}\psi_1\otimes
  \Omega}\\
  & = & e_1 + \alpha^3 \sprod{\psi_1\otimes\Omega}{
  A(\alpha x)^2\psi_1\otimes\Omega}\\
  &\leq & e_1 + 4 c_2(\kappa) \alpha^3
\end{eqnarray*}
by Lemma~\ref{lm:basic}. To prove Statement 2 we first estimate
$H_{\sigma}$ from below in terms of $\Hpart$. By Lemma~\ref{lm:basic},
\begin{eqnarray*}
   H_{\sigma}  &=& \Hpart + 2 p\cdot A(\alpha x) \alpha^{3/2} + A(\alpha x)^2\alpha^3 + H_f\\
   &\geq & \Hpart -\frac{8}{3}c_1(\kappa) \alpha^3 p^2.
\end{eqnarray*}
Since $p^2\leq 3(\Hpart + D)$ for some constant $D$, it follows that
\[
   H_{\sigma} \geq \Hpart(1-8c_1(\kappa)\alpha^3) - 8c_1(\kappa)D\alpha^3.
\]
By Perrson's theorem, $\sprod{\ph}{(\Hpart\otimes 1)\ph}\geq e_2-o_R(1)$, as $R\to
\infty$, for normalized $\ph\in D_R$, with $\|\ph\|=1$, and by assumption $1- 8c_1(\kappa)\alpha^3\geq 0$. Hence we obtain
\begin{eqnarray*}
   \Sigma_{R,\sigma} &\geq & (e_2-o_R(1))(1-8c_1(\kappa)\alpha^3)-
   8c_1(\kappa)D\alpha^3\\
   &=& e_2 - o_R(1)(1-8c_1(\kappa)\alpha^3) - 8c_1(\kappa)\alpha^3 (e_2+D),
\end{eqnarray*}
which proves the lemma.
\end{proof}

%%%%%%%%%%%%%%%%%%%%%%%%%%%%%%%%%%%%%%%%%%%%%%%%%%%%%%%%%%%%%%%

\begin{lemma}[Electron localization]\label{qed-x-loc}
  For every $\lambda<e_2$ there exists $\alpha_{\lambda}>0$ such that for all
  $\alpha\leq \alpha_{\lambda}$ and all $n\in \N$
\[
     \sup_{\sigma\geq 0}\||x|^{n}E_{\lambda}(H_{\sigma})\|<\infty.
\]
\end{lemma}

\begin{proof}
From \cite[Theorem 1]{Griesemer2004} we know that
\(\|e^{\eps|x|} E_{\lambda}(H_{\sigma})\| <\infty \)
if $\lambda+\eps^2 < \Sigma_{\sigma}$. Moreover, from the proof of that
theorem we see that
\begin{equation*}
  \sup_{\sigma\geq 0}\|e^{\eps|x|}E_{\lambda}(H_{\sigma})\| <\infty
\end{equation*}
if $R>0$ and $\delta>0$ can be found so that
\begin{equation}
  \label{qed-loc}
  \Sigma_{\sigma,R}-\frac{\tilde{C}}{R^2} \geq \lambda +\eps^2 +\delta
\end{equation}
holds uniformly in $\sigma$. Here $\tilde{C}$ is a constant that is independent
of the system. Given $\lambda<e_2$, pick $\alpha_{\lambda}>0$ so small that
$e_2-c_1(\kappa)\alpha_{\lambda}^3C>\lambda$ with $C$ as in Lemma~\ref{lm:E-sigma}. It then follows from
Lemma~\ref{lm:E-sigma} that \eqref{qed-loc} holds true for some $\delta>0$ if $R$ is
large enough.
\end{proof}

%%%%%%%%%%%%%%%%%%%%%%%%%%%%%%%%%%%%%%%%%%%%%%%%%%%%%%%%%%%%%

\begin{theorem}[Spectral gap]\label{thm:gap}
If $\alpha\ll 1$  then
$$
   \sigma(H_{\sigma}\upharpoonright\H^{\sigma})\cap (E_{\sigma},E_{\sigma}+\sigma) =
   \emptyset
$$
for all $\sigma\leq (e_2-e_1)/2$.
\end{theorem}

\noindent\emph{Remark.} Variants of this results are already known \cite{FGS3,
  BFP1}. 

\begin{proof}
From \cite{GLL} we know that
$$
   \inf\sigma_{ess}(H_{\sigma}\upharpoonright\H^{\sigma}) \geq
   \min(E_{\sigma}+\sigma,\Sigma_{\sigma}).
$$
On the other hand, by Lemma~\ref{lm:E-sigma},
$$
   \Sigma_{\sigma}-E_{\sigma} \geq e_2-e_1 -\alpha^3(8c_1(\kappa)+4c_2(\kappa)) \geq \sigma
$$
under our assumptions on $\alpha$ and $\sigma$. This proves that
$$
   \inf\sigma_{ess}(H_{\sigma}\upharpoonright\H^{\sigma}) \geq E_{\sigma}+\sigma.
$$
From Proposition~\ref{prop:pos-comm}, below, it follows that $H_{\sigma}$ has no eigenvalues in
$(E_{\sigma},E_{\sigma}+\sigma)$.
\end{proof}

In order to complete the proof of Theorem~\ref{thm:gap}, we need a further
commutator estimate and a corresponding Virial Theorem. We define $\tilde{B}=\dGamma(\hat{b})+
\alpha^{3/2}x\cdot\phi(i\hat{b}\tilde{\chi}^{\sigma}G_0)$ where
$\hat{b}=(\hat{k}\cdot y+ y\cdot \hat{k})/2$ and $\hat{k}=k/|k|$, and begin
with a formal computation of the commutator $[H_{\sigma},i\tilde{B}]$. To this
end we set  $\Pi_{\sigma}=p+\alpha^{3/2}A^{\sigma}(\alpha x)$ so that
$H_{\sigma}=\Pi_{\sigma}^2+V+H_f$. It follows that 
\begin{equation*}
   [H_{\sigma}, i\tilde{B}] = \Pi_{\sigma}[\Pi_{\sigma},i\tilde{B}] + [\Pi_{\sigma},i\tilde{B}]\Pi_{\sigma}+[H_f,i\tilde{B}]
\end{equation*}
where
\begin{equation*}
[H_f,i\tilde{B}] = N- \alpha^{3/2}x\cdot\phi(\omega\hat{b}\tilde{\chi}^{\sigma}G_0)
\end{equation*}
and
\begin{eqnarray}\nonumber
  [\Pi_{\sigma},i\tilde{B}] &=& \big[\Pi_{\sigma},i\dGamma(\hat{b})\big]
  +\big[\Pi_{\sigma}, i\alpha^{3/2} x\cdot\phi(i\hat{b}\tilde{\chi}^{\sigma}G_0)\big]\\ \nonumber
  &=& -\alpha^{3/2}\phi(i\hat{b}\tilde{\chi}^{\sigma}G_x) +
  \alpha^{3/2}\phi(i\hat{b}\tilde{\chi}^{\sigma}G_0)
  -2\alpha^3\Rea\sprod{\tilde{\chi}^{\sigma}G_x}{x\hat{b}\tilde{\chi}^{\sigma}G_0}\\ \label{pos-com1}
   &=&  -\alpha^{3/2}\phi(i\hat{b}\tilde{\chi}^{\sigma}\Delta G_x) -2\alpha^3\Rea\sprod{\tilde{\chi}^{\sigma}G_x}{x\hat{b}\tilde{\chi}^{\sigma}G_0}.
\end{eqnarray}
Here $\Delta G_{x}=G_{x}-G_0$. The resulting expression for $[H_{\sigma},i\tilde{B}]$ is our \emph{definition}
of this commutator as a quadratic form on $\ran E_{(0,\sigma)}(H_{\sigma}-E_{\sigma})$,
where  $\alpha \ll 1$ and $0<\sigma\leq \egap/2$ are assumed. The reason for the contribution
$\alpha^{3/2}x\cdot\phi(i\hat{b}\tilde{\chi}^{\sigma}G_0)$ to the
operator $\tilde{B}$ is that in Equation~\eqref{pos-com1} it leads
to $\phi(i\hat{b}\tilde{\chi}^{\sigma}\Delta G_{x})$ rather than
$\phi(i\hat{b}\tilde{\chi}^{\sigma}G_{x})$. The more regular behavior of
$\Delta G_{x}(k)$ as $k\to 0$ is essential to get estimates that hold
\emph{uniformly} in $\sigma\in(0,\egap/2)$. 

The following proposition completes the proof of Theorem~\ref{thm:gap}.

\begin{prop}\label{prop:pos-comm}
Let $[H_{\sigma},i\tilde{B}]$ be defined as above and suppose that 
$\alpha \ll 1$ and $0<\sigma\leq \egap/2$. Then
$$
    E_{(0,\sigma)}(H_{\sigma}-E_{\sigma})[H_{\sigma},i\tilde{B}]E_{(0,\sigma)}(H_{\sigma}-E_{\sigma})
    \geq \frac{1}{2}E_{(0,\sigma)}(H_{\sigma}-E_{\sigma}),
$$
and moreover, if $H_{\sigma}\ph=E\ph$ with $E-E_{\sigma}\in(0,\sigma)$, then
$\sprod{\ph}{[H_{\sigma},i\tilde{B}]\ph}=0$.
\end{prop}

\begin{proof}
We first show that $[H_{\sigma},i\tilde{B}]-N$ between spectral projections
$E_{(0,\sigma)}(H_{\sigma}-E_{\sigma})$ is $O(\alpha^{3/2})$ as
$\alpha\to 0$. To this end we set $\lambda=(1/4)e_1+(3/4)e_2$ and prove
Steps 1-3 below. Note that, by Lemma~\ref{lm:E-sigma}, $E_{\sigma}+\sigma\leq \lambda$
for $\sigma\leq \egap/2$ and $2c_{2}(\kappa)\alpha^3\leq \egap/4$.

\noindent\underline{Step 1}.
$$
  \sup_{\sigma>0}\|E_{\lambda}(H_{\sigma})x\cdot\phi(\omega\hat{b}\tilde{\chi}^{\sigma}G_0)E_{\lambda}(H_{\sigma})\|<\infty.
$$
One has the estimate
$$
  \|E_{\lambda}(H_{\sigma})x\cdot\phi(\omega\hat{b}\tilde{\chi}^{\sigma}G_0)E_{\lambda}(H_{\sigma})\|
  \leq \|E_{\lambda}(H_{\sigma})x\|
  \|\omega\hat{b}\tilde{\chi}^{\sigma}G_0\|_{\omega}\|(H_f+1)^{1/2}E_{\lambda}(H_{\sigma})\|
$$
where each factor is bounded uniformly in $\sigma>0$. For
the first one this follows from Lemma~\ref{qed-x-loc}, for the
second one from
$|\omega\hat{b}\tilde{\chi}^{\sigma}G_0(k)|=O(|k|^{-1/2})$ and for
the third one from
$\sup_{\sigma}\|(H_f+1)^{1/2}(H_{\sigma}+1)^{-1}\|<\infty$, by Lemma~\ref{lm:basic}.

\noindent\underline{Step 2}.
$$
\sup_{\sigma>0}\|E_{\lambda}(H_{\sigma})\Pi_{\sigma}\cdot
\phi(i\hat{b}\tilde{\chi}^{\sigma}\Delta G_x)E_{\lambda}(H_{\sigma})\| <\infty.
$$
This time we use
\begin{eqnarray}\nonumber
  \lefteqn{\|E_{\lambda}(H_{\sigma})\Pi_{\sigma}\cdot
\phi(i\hat{b}\tilde{\chi}^{\sigma}\Delta G_x)
E_{\lambda}(H_{\sigma})\|}\\ &\leq & \|E_{\lambda}(H_{\sigma})\Pi_{\sigma}\|
\Big(\sup_{x}\expect{x}^{-1}\|\hat{b}\tilde{\chi}^{\sigma}\Delta G_x\|_{\omega}\Big)
\|\expect{x}(H_f+1)^{1/2}E_{\lambda}(H_{\sigma})\|.\label{eq:S2.1}
\end{eqnarray}
Since
\begin{eqnarray*}
  \hat{b}\tilde{\chi}^{\sigma}\Delta G_x(k,\lambda) &=& i\left(\partial_{|k|}+|k|^{-1}\right)
  \tilde{\chi}^{\sigma}(e^{-ik\cdot x} -1)\frac{\kappa(k)}{\sqrt{|k|}}\eps_{\lambda}(k)\\
  &=& O(\expect{x}|k|^{-1/2}), \qquad (k\to 0),
\end{eqnarray*}
while, as $k\to\infty$, it decays like a Schwartz-function, it follows that
$$
  \sup_{x,\sigma}\expect{x}^{-1}\|\hat{b}\tilde{\chi}^{\sigma}\Delta G_x\|_{\omega} <\infty.
$$
The first factor of \eqref{eq:S2.1} is bounded uniformly in $\sigma>0$ thanks
to Lemma~\ref{lm:basic}, and for the last one we
have
$$
  \|\expect{x}(H_f+1)^{1/2}E_{\lambda}(H_{\sigma})\|\leq \|\expect{x}^2E_{\lambda}(H_{\sigma})\|+\|(H_f+1)E_{\lambda}(H_{\sigma})\|,
$$
which, by Lemma~\ref{qed-x-loc} and Lemma~\ref{lm:basic}, is also bounded uniformly in $\sigma$.

\noindent\underline{Step 3}.
$$
  \sup_{\sigma}\|E_{\lambda}(H_{\sigma})\Pi_{\sigma}\cdot
 \Rea \sprod{\tilde{\chi}^{\sigma}G_x}{x\cdot
 \hat{b}\tilde{\chi}^{\sigma}G_0}E_{\lambda}(H_{\sigma})\| <\infty.
$$
This follows from estimates in the proof of Step 2.

From Steps 1, 2, 3 and $N\geq 1-P_{\Omega}$ it follows that
\begin{equation}\label{eq:S3-4}
   E_{\lambda}(H_{\sigma})[H_{\sigma},i\tilde{B}] E_{\lambda}(H_{\sigma}) \geq
   E_{\lambda}(H_{\sigma})(1-P_{\Omega})E_{\lambda}(H_{\sigma})+O(\alpha^{3/2}).
\end{equation}
In Steps 4, 5, and 6 below we will show that
$E_{(0,\sigma)}(H_{\sigma}-E_{\sigma})P_{\Omega}E_{(0,\sigma)}(H_{\sigma}-E_{\sigma})=O(\alpha^{3/2})$
as well. Hence the proposition will follow from \eqref{eq:S3-4}.

Let $\Ppart$ be the ground state projection of $-\Delta+V$ and let
$\Ppart^{\perp}=1-\Ppart$. Recall that $\Ppart$ is a projection of rank one,
by assumption on $e_1=\inf\sigma(-\Delta+V)$. 

\noindent\underline{Step 4}.
$$
   \|(\Ppart^{\perp}\otimes P_{\Omega}) E_{\lambda}(H_{\sigma})\| = O(\alpha^{3/2}).
$$
Let $H^{(0)}$ denote the Hamiltonian $H$ with $\alpha=0$ and let
$f\in C_0^{\infty}(\R)$ with $\supp(f)\subset(-\infty,e_2)$ and
$f=1$ on $[\inf_{\sigma\leq\egap}E_{\sigma},\lambda]$. Then
$E_{\lambda}(H_{\sigma}) = f(H_{\sigma})E_{\lambda}(H_{\sigma})$,
$(\Ppart^{\perp}\otimes P_{\Omega})f(H^{(0)})=0$ and
$$
  f(H_{\sigma})-f(H^{(0)}) = \int \rd \tilde{f}(z)
  \frac{1}{z-H_{\sigma}}\big(2\alpha^{3/2}p\cdot
  A^{\sigma}(\alpha x)+\alpha^3 A^{\sigma}(\alpha x)^2\big)\frac{1}{z-H^{(0)}}=
  O(\alpha^{3/2}).
$$
It follows that
\begin{eqnarray*}
  \|(\Ppart^{\perp}\otimes P_{\Omega}) E_{\lambda}(H_{\sigma})\| &=&
  \|(\Ppart^{\perp}\otimes
  P_{\Omega})\big[f(H_{\sigma})-f(H^{(0)})\big]
  E_{\lambda}(H_{\sigma})\|\\
  &\leq & \|f(H_{\sigma})-f(H^{(0)})\| = O(\alpha^{3/2}).
\end{eqnarray*}

\noindent\underline{Step 5}. Let $P_{\sigma}$ denote the ground
state projection of $H_{\sigma}$. Then
$$
  \|\Ppart\otimes P_{\Omega} - P_{\sigma}\|= O(\alpha^{3/2}).
$$
Since $1-P_{\Omega}\leq N^{1/2}$ we have
\begin{eqnarray*}
    1-\Ppart\otimes P_{\Omega} &=&  1-P_{\Omega} + \Ppart^{\perp}\otimes P_{\Omega}\\
      &\leq & N^{1/2} + \Ppart^{\perp}\otimes P_{\Omega}
\end{eqnarray*}
where $\|(\Ppart^{\perp}\otimes P_{\Omega})P_{\sigma}\| =
O(\alpha^{3/2})$ by Step~4 and
$\|N^{1/2}P_{\sigma}\|=O(\alpha^{3/2})$ by Lemma~\ref{lm6}. It follows that
$\|(1-\Ppart\otimes P_{\Omega})P_{\sigma}\|= O(\alpha^{3/2})$.
Hence, for $\alpha$ small enough, $P_{\sigma}$ is of
rank one and the assertion of Step 5 follows.

\noindent\underline{Step 6}.
$$
    E_{(0,\sigma)}(H_{\sigma}-E_{\sigma})(1\otimes P_{\Omega})E_{(0,\sigma)}(H_{\sigma}-E_{\sigma}) = O(\alpha^{3/2}).
$$
Since $P_{\sigma}E_{(0,\sigma)}(H_{\sigma}-E_{\sigma})=0$, it follows from
Step 4 and Step 5 that
\begin{eqnarray*}
   \lefteqn{\|(1\otimes P_{\Omega}) E_{(0,\sigma)}(H_{\sigma}-E_{\sigma})\| =
   \|(1\otimes P_{\Omega}-P_{\sigma})E_{(0,\sigma)}(H_{\sigma}-E_{\sigma})\|}\\
  &\leq &  \|(\Ppart\otimes P_{\Omega}-P_{\sigma})E_{(0,\sigma)}(H_{\sigma}-E_{\sigma})\|+ \|(\Ppart^{\perp}\otimes P_{\Omega}) E_{(0,\sigma)}(H_{\sigma}-E_{\sigma})\|\\
  & = &  O(\alpha^{3/2}).
\end{eqnarray*}

In order to prove the Virial Theorem, $\sprod{\ph}{[H_{\sigma},i\tilde{B}]\ph}=0$,
for eigenvectors $\ph$ with energy $E\in(E_{\sigma},E_{\sigma}+\sigma)$ we
approximate $\tilde{B}$ with suitably regularized operators $\tilde{B}_{\eps}$, $\eps>0$, that are defined on
$D(H_{\sigma})$, and converge to $\tilde{B}$ as $\eps\to 0$, in the sense that 
$[H_{\sigma},i\tilde{B}_{\eps}]\to [H_{\sigma},i\tilde{B}]$ weakly as $\eps\to
0$. The Virial Theorem for $[H_{\sigma},i\tilde{B}_{\eps}]$ then implies the
asserted Virial Theorem. The infrared cutoff $\sigma$ is crucial for this to
work. For more details, see, e.g., \cite{FGS2}, Appendix E. 
\end{proof}

%%%%%%%%%%%%%%%%%%%%%%%%%%%%%%%%%%%%%%%%%%%%%%%%%%%%%%

%%%%%%%%%%%%%%%%%%%%%%%  ground state photons  %%%%%%%%%%%%%%%%%%%%%%%%%%%%%%%%%%

\begin{lemma}[ground state photons]\label{lm6}
Suppose $H_{\sigma}P_{\sigma}=E_{\sigma}P_{\sigma}$ where $\sigma\geq 0$,
$E_{\sigma}=\inf\sigma(H_{\sigma})$, and $P_{\sigma}$ is the ground state
projection of $H_{\sigma}$. Here $H_{\sigma=0}=H$. Let $R_{\sigma}(\omega)=(H_{\sigma}-E_{\sigma}+\omega)^{-1}$. Then
\begin{eqnarray*}
 (i)\qquad a(k)P_{\sigma} &=& -i\alpha^{3/2}\Big[1-\omega R_{\sigma}(\omega)
 -2R_{\sigma}(\omega)(\Pi_{\sigma}\cdot k)+\alpha
  R_{\sigma}(\omega)k^2\Big]x\cdot G_{x}(k)^{*}P_{\sigma}\\
   && -2\alpha^{3/2}R_{\sigma}(\omega) k\cdot G_{\alpha
  x}(k)^{*}P_{\sigma}.
\end{eqnarray*}
There are constants $C,D$ independent of $\sigma,\alpha\in [0,1]$
such that
\begin{equation*}
\begin{aligned}
  (ii)\quad &\|a(k)P_{\sigma}\| \leq  \alpha^{3/2}\frac{C}{|k|^{1/2}},\\
  (iii)\quad &\|x a(k)P_{\sigma}\| \leq  \alpha^{3/2}\frac{D}{|k|^{3/2}}.
\end{aligned}
\end{equation*}
\end{lemma}

\begin{proof}
We suppress the subindex $\sigma$ for notational simplicity. By
the usual pull-through trick
\begin{eqnarray*}
(H-E+\omega(k))a(k)P &=& [H,a(k)]\ph +\omega(k)a(k)P\\
&=& -\alpha^{3/2} 2\Pi\cdot G_{x}(k)^{*}P.
\end{eqnarray*}
Since $2\Pi=i[H,x]=i[H-E,x]$, and $(H-E)\ph=0$ we can rewrite this
as
\begin{eqnarray}
  i\alpha^{-3/2}a(k)\ph &=& R(\omega)\big[(H-E)x-x(H-E)\big]G_{\alpha
  x}(k)^{*}P\nonumber \\
  &=& (1-\omega R(\omega))(x\cdot G_{x}(k)^{*})P- R(\omega) x[H,G_{\alpha
  x}(k)^{*}]P\label{lm6.1}
\end{eqnarray}
For the commutator we get
\begin{eqnarray*}
   [H,G_{x}(k)^{*}] &=& (\Pi\cdot k)G_{x}(k)^{*} + G_{\alpha
   x}(k)^{*}(\Pi\cdot k)\\
   &=& 2(\Pi\cdot k)G_{x}(k)^{*}-\alpha k^2 G_{x}(k)^{*}
\end{eqnarray*}
and hence, using $x(\Pi\cdot k)=(\Pi\cdot k)x+ik$,
\begin{equation}\label{lm6.2}
  x[H,G_{x}(k)^{*}] = \big[2(\Pi\cdot k)-\alpha k^2\big]x\cdot G_{\alpha
  x}(k)^{*}+ 2ik\cdot G_{x}(k)^{*}.
\end{equation}
From \eqref{lm6.1} and \eqref{lm6.2} we conclude that
\begin{eqnarray*}
  i\alpha^{-3/2}a(k)P &=&\big[1-\omega R(\omega)-2R(\omega)(\Pi\cdot k)+\alpha R(\omega)k^2\big]
  x\cdot G_{x}(k)^{*}P\\
  & &-2iR(\omega) k\cdot G_{x}(k)^{*}P.
\end{eqnarray*}
(ii) First of all $\sup_{\sigma\geq 0}\|x P\|<\infty$ by
Lemma~\ref{qed-x-loc} and $|G_{x}(k)|\leq \const|k|^{-1/2}$ by
definition of $G_x(k)$. Since $\|R(\omega)\|\leq |k|^{-1}$ and
$\|R(\omega)\Pi\|\leq \const(1+|k|^{-1})$ we find that
\[
   \Big\|\Big[1-\omega R(\omega)-2R(\omega)(\Pi\cdot k)+\alpha
  R(\omega)k^2\Big]\Big\| \leq \const\qquad \text{for}\ \alpha, |k|\leq 1
\]
This proves (ii). To estimate the norm of $xa(k)P$ we
use (i) and commute $x$ with all operators in front of
$P$ so that we can apply Lemma~\ref{qed-x-loc} to the
operator $x^2P$. Since
\[
   [x,R(\omega)] = -2i R(\omega) \Pi R(\omega)
\]
the resulting estimate for $\|xa(k)P\|$ is worse by one power of
$|k|$ than our estimate (i) for $\|a(k)P\|$.
\end{proof}

%%%%%%%%%%%%%%%%%%%%%%%%%%% overlap %%%%%%%%%%%%%%%%%%%%%%%%%%% 
%%%%%%%%%%%%%%%%%%%%%%%%%%%%%%%%%%%%%%%%%%%%

The following two Lemmas are consequences of Lemma~\ref{lm6}.

\begin{lemma}[overlap estimate]\label{lm:overlap}
Let $P^{\sigma}\otimes f_{\Delta}(H_{f,\sigma})$ on
$\H^{\sigma}\otimes\F_{\sigma}$ and $\chi_{\sigma}$
be defined as in Section~\ref{sec:qed-mourre}. For every $\mu>-1$ there exists
a constant $C_{\mu}$, such that for all $\alpha\in [0,1]$, for all $\sigma\in[0,\egap/2]$ and for every
function $h_{x}\in L^2(\R^3)$, depending parametrically on the electron
position $x\in\R^3$, with $|h_{x}(k)|\leq |k|^{\mu}\expect{x}$,
$$
    \|a(\chi_{\sigma}h_x)P^{\sigma}\otimes f_{\Delta}(H_{f,\sigma})\| 
\leq C_{\mu}\sigma^{\mu+3/2}\|\expect{x}P^{\sigma}\|.
$$
Here $\expect{x}=\sqrt{1+x^2}$.
\end{lemma}

\begin{proof}
Let $\ph\in \H^{\sigma}\otimes\F_{\sigma}$ with $\|\ph\|=1$. By
construction of $\chi_{\sigma}$,
\begin{eqnarray*}
  a(\chi_{\sigma}h_x)P^{\sigma}\otimes f_{\Delta}(H_{f,\sigma})\ph &=&
  \int_{\sigma\leq |k|\leq 2\sigma}\chi_{\sigma}(k)\overline{h_x(k)}a(k) P^{\sigma}\otimes
  f(H_{f,\sigma})\ph \, dk\\
  & & +\int_{|k|<\sigma}\chi_{\sigma}(k)\frac{\overline{h_x(k)}}{|k|^{1/2}}P^{\sigma}\otimes
  |k|^{1/2}a(k)f(H_{f,\sigma})\ph\, dk.\\
\end{eqnarray*}
Using $|\chi_{\sigma}h_x(k)|\leq |k|^{\mu}\expect{x}$,
$\|f_{\Delta}(H_{f,\sigma})\|\leq 1$, and the Cauchy-Schwarz
inequality applied to the second integral we obtain
\begin{eqnarray*}
  \lefteqn{\|a(\chi_{\sigma} h_x)P^{\sigma}\otimes
  f_{\Delta}(H_{f,\sigma})\ph\|}\\
   & \leq & \int_{\sigma\leq |k|\leq 2\sigma} |k|^{\mu}
   \|\expect{x}a(k)P^{\sigma}\|\,  dk  + \left(\int_{|k|\leq \sigma}|k|^{2\mu-1}\, dk\right)^{1/2}
   \|\expect{x}P^{\sigma}\|\,\|H_{f,\sigma}^{1/2}f(H_{f,\sigma})\ph\|.
\end{eqnarray*}
The Lemma now follows from  Lemma~\ref{lm6} and $\|H_{f,\sigma}^{1/2}f(H_{f,\sigma})\| \leq \sigma^{1/2}$.
\end{proof}

%%%%%%%%%%%%%%%%%%%%%   E - E_sigma  %%%%%%%%%%%%%%%%%%%%%%%%%%%%%%

\begin{lemma}\label{lm5}
There exists a constant $C$ such that
$$
  |E-E_{\sigma}| = C \alpha^{3/2} \sigma^2
$$
for all $\sigma\geq 0$ and $\alpha\in [0,1]$.
\end{lemma}

\begin{proof}
Let $\psi$ and $\psi_{\sigma}$ be normalized ground states of $H$
and $H_{\sigma}$ respectively. Then, by Rayleigh-Ritz,
\begin{align}\label{ev1}
  E-E_{\sigma} &\leq
  \sprod{\psi_{\sigma}}{(H-H_{\sigma})\psi_{\sigma}}\\ \label{ev2}
  E_{\sigma}-E &\leq
  \sprod{\psi}{(H_{\sigma}-H)\psi}
\end{align}
where $H-H_{\sigma} = \Pi^2-\Pi_{\sigma}^2$ and
\begin{eqnarray}\nonumber
   \Pi^2-\Pi_{\sigma}^2 &=& 2\alpha^{3/2} p\cdot
   (A(\alpha x)-A^{\sigma}(\alpha x))\\
   && + \alpha^3[A(\alpha x)+A^{\sigma}(\alpha x)]\cdot[A(\alpha
   x)-A^{\sigma}(\alpha x)].\label{ev4}
\end{eqnarray}
To estimate the contribution due to \eqref{ev4} we note that
\begin{eqnarray}
\nonumber
  [A(\alpha x)+A^{\sigma}(\alpha x)]\cdot[A(\alpha x)-A^{\sigma}(\alpha x)]&=&
  [A(\alpha x)+A^{\sigma}(\alpha x)]\cdot a(\chi_{\sigma} G_{x})\\
  & &+ a^{*}(\chi_{\sigma} G_{x})\cdot [A(\alpha x)+A^{\sigma}(\alpha x)]\label{ev5}\\
  & &+ 2\int |G_{x}(k)|^2\chi_{\sigma}^2\, dk.\nonumber
\end{eqnarray}
The last term in \eqref{ev5} is of order $\sigma^2$ and from
Lemma~\ref{lm6} it follows that
\begin{equation}
\|a(\chi_{\sigma}G_{x})\psi_{\sigma}\|,\
\|a(\chi_{\sigma}G_{x})\psi\| \leq C\alpha^{3/2} \int_{|k|\leq
2\sigma}|G_{x}(k)|\frac{1}{\sqrt{|k|}} dk=
O(\alpha^{3/2}\sigma^2)
\end{equation}
Moreover, by Lemma~\ref{lm:basic},
\begin{equation*}
  \|p\psi_{\sigma}\|,
  \|[A(\alpha x)+A_{\sigma}(\alpha x)]\psi_{\sigma}\|\leq \const.
\end{equation*}
It follows that the contributions of \eqref{ev4} to \eqref{ev1}
and \eqref{ev2} are of order $\alpha^{3/2}\sigma^2$ and
$\alpha^3\sigma^2$.
\end{proof}

%%%%%%%%%%%%%%%%%%%%%%%%%%%%%%%%%%%%%%%%%%%%%%%%%%%%%%%%%%%%%%%%%%%%%%%%%
\section{Conjugate Operator Method}
\label{sec:conjugate}

In this section we describe the conjugate operator method in the
version of Amrein, Boutet de Monvel, Georgescu, and Sahbani
\cite{Amreinetal1996, Sahbani1997}. In the paper of Sahbani the
theory of Amrein et al. is generalized in a way that is crucial
for our paper. For simplicity, we present a weaker form of the
results of Sahbani with comparatively stronger assumptions that
are satisfied by our Hamiltonians.

The conjugate operator method to analyze the spectrum of a
self-adjoint operator $H: D(H)\subset\H\to\H$ assumes the
existence of another self-adjoint operator $A$ on $\H$, the
conjugate operator, with certain properties. The results below
yield  information on the spectrum of $H$ in an open subset
$\Omega\subset \R$, provided the following assumptions hold:

\begin{itemize}
\item[(i)] \emph{$H$ is locally of class $C^2(A)$ in $\Omega$.} This
  assumption means that the mapping
\[
   s\mapsto e^{-iAs}f(H)e^{iAs}\ph
\]
is twice continuously differentiable, for all $f\in C_0^{\infty}(\Omega)$ and
all $\ph\in \H$.

\item[(ii)] For every $\lambda\in \Omega$, there exists a neighborhood
  $\Delta$ of $\lambda$ with $\overline{\Delta}\subset \Omega$, and a constant $a>0$
  such that
\[
    E_{\Delta}(H)[H,iA]E_{\Delta}(H)\geq a E_{\Delta}(H).
\]
\end{itemize}

\emph{Remarks:} By (i), the commutator $[H,iA]$ is well defined as
a sesquilinear form on the intersection of $D(A)$ and $\cup_{K}
E_{K}(H)\H$, where the union is taken over all compact subsets $K$
of $\Omega$. By continuity it can be extended to $\cup_{K}
E_{K}(H)\H$.

The following two theorems follow from Theorems~0.1 and 0.2 in
\cite{Sahbani1997} and assumptions (i) and (ii), above.

\begin{theorem}\label{thm:LAP}
For all $s>1/2$ and all $\ph,\psi\in \H$, the limit
\[
   \lim_{\eps\to 0+}\sprod{\ph}{\expect{A}^{-s}R(\lambda\pm i\eps)\expect{A}^{-s}\psi}
\]
exists uniformly for $\lambda$ in any compact subset of $\Omega$. In
particular, the spectrum of $H$ is purely absolutely continuous in $\Omega$.
\end{theorem}

This theorem allows one to define operators $\expect{A}^{-s}R(\lambda\pm
i0)\expect{A}^{-s}$ in terms of the sesquilinear forms
$$
   \sprod{\ph}{\expect{A}^{-s}R(\lambda\pm i0)\expect{A}^{-s}\psi}
   = \lim_{\eps\to 0}\sprod{\ph}{\expect{A}^{-s}R(\lambda\pm i\eps)\expect{A}^{-s}\psi}.
$$
By the uniform boundedness principle these operators are bounded.

\begin{theorem}\label{thm:Holder}
If $1/2<s<1$ then
\[
   \lambda\mapsto \expect{A}^{-s}R(\lambda\pm i0)\expect{A}^{-s}
\]
is locally H\"older continuous of degree $s-1/2$ in $\Omega$.
\end{theorem}

\begin{theorem}\label{thm:decay-in-time}
Suppose assumptions (i) and (ii) above are satisfied,
$s\in (1/2,1)$, and $f\in C_0^{\infty}(\Omega)$. Then
\begin{equation*}
  \| \expect{A}^{-s}e^{-iHt}f(H)\expect{A}^{-s}\|
  = O\left(\frac{1}{t^{s-1/2}}\right), \qquad (t\to\infty).
\end{equation*}
\end{theorem}

\begin{proof}
For every $f\in C_0^{\infty}(\R)$ and all $\ph\in \H$
\begin{equation}\label{eq:pw1}
  e^{-iHt}f(H)\ph = \lim_{\eps\downarrow 0}\frac{1}{\pi}\int e^{-i\lambda
  t}f(\lambda)\Ima(H-\lambda-i\eps)^{-1}\ph\, d\lambda
\end{equation}
by the spectral theorem. Now suppose $f\in C_0^{\infty}(\Omega)$
and set
$F(z)=\pi^{-1}\expect{A}^{-s}\Ima(H-z)^{-1}\expect{A}^{-s}$. Then
\eqref{eq:pw1} and Theorem~\ref{thm:LAP} imply
\begin{equation}\label{eq:pw2}
  \expect{A}^{-s}e^{-iHt}f(H)\expect{A}^{-s}\ph=
  \int e^{-i\lambda
  t}f(\lambda)F(\lambda+i0)\ph\, d\lambda
\end{equation}
In this equation we replace $H$ by $H-\pi/t$ with $t$ so large that
$f(\cdot-\pi/t)$ has support in $\Omega$. Then it becomes
\begin{equation}\label{eq:pw3}
  \expect{A}^{-s}e^{-iHt}f(H-\pi/t)\expect{A}^{-s}\ph=
  -\int e^{-i\lambda t}f(\lambda)F(\lambda+\pi/t+i0)\ph\,
  d\lambda.
\end{equation}
Taking the sum of \eqref{eq:pw2} and \eqref{eq:pw3} and using
$\|f(H)-f(H-\pi/t)\|=O(t^{-1})$, which may be derived from the almost analytic
functional calculus, see \eqref{HS-calculus}, we get
\begin{eqnarray*}
2\lefteqn{ \|\expect{A}^{-s}e^{-iHt}f(H)\expect{A}^{-s}\|
+ O(t^{-1})}\\
 &\leq & \int|f(\lambda)|\|F(\lambda+i0)-F(\lambda+\pi/t+i0)\|
d\lambda = O(1/t^{s-1/2}),
\end{eqnarray*}
where the H\"older continuity from Theorem~\ref{thm:Holder} was
used in the last step.
\end{proof}

For completeness we also include the Virial Theorem (Proposition~3.2 of \cite{Sahbani1997}):

\begin{prop}\label{Virial}
If $\lambda\in \Omega$ is an eigenvalue of $H$ and $E_{\{\lambda\}}(H)$ denotes
the projection onto the corresponding eigenspace, then
$$
   E_{\{\lambda\}}(H)[H,iA]E_{\{\lambda\}}(H) = 0.
$$
\end{prop}

In the remainder of this section we introduce tools that will help us to
verify assumption (i). To begin with we recall, from
\cite{Amreinetal1996,Sahbani1997}, that a bounded operator $T$ on $\H$ is said to be
of class $C^k(A)$ if the mapping
$$
   s\mapsto e^{-iAs}Te^{iAs}\ph
$$
is $k$ times continuously differentiable for every $\ph\in \H$. The following propositions
summarize results in Lemma~6.2.9 and Lemma~6.2.3 of \cite{Amreinetal1996}.

\begin{prop}\label{C1-tool}
Let $T$ be a bounded operator on $\H$ and let $A=A^*:
D(A)\subset\H\to\H$. Then the following are equivalent.
\begin{itemize}
\item[(i)] $T$ is of class $C^{1}(A)$.
\item[(ii)] There is a constant $c$ such that for all $\ph,\psi\in D(A)$
\[
    |\sprod{A\ph}{T\psi}-\sprod{\ph}{TA\psi}| \leq c\|\ph\|
    \|\psi\|.
\]
\item[(iii)] $\liminf_{s\to
0+}\frac{1}{s}\big\|e^{-iAs}Te^{iAs}-T\big\|<\infty$.
\end{itemize}
\end{prop}

\begin{proof}
If $T$ is of class $C^{1}(A)$ then $\sup_{s\neq 0}\|s^{-1}(e^{-iAs}Te^{iAs}-T)\|<\infty$
 by the uniform boundedness principle. Thus statement (i) implies statement
 (iii). To prove the remaining assertions we use that, for all $\ph,\psi\in D(A)$,
\begin{equation}\label{c1a-1}
  \frac{1}{s}\sprod{\ph}{(e^{-iAs}Te^{iAs}-T)\psi} =
  \frac{-i}{s}\int_0^s d\tau\Big[\sprod{e^{iA\tau}A\ph}{T e^{iA\tau}\psi}
  -\sprod{e^{iA\tau}\ph}{Te^{iA\tau}A\psi}\Big].
\end{equation}
Since the integrand is a continuous function of $\tau$, its value at $\tau=0$,
$\sprod{A\ph}{T\psi}-\sprod{\ph}{TA\psi}$, is the limit of \eqref{c1a-1} as
$s\to 0$. It follows that
\begin{equation}\label{c1a-2}
\begin{split}
  |\sprod{A\ph}{T\psi}-\sprod{\ph}{TA\psi}| &= \lim_{s\to
  0+}s^{-1}|\sprod{\ph}{(e^{-iAs}Te^{iAs}-T)\psi}|\\
  &\leq \liminf_{s\to 0+} s^{-1}\|e^{-iAs}Te^{iAs}-T\|
  \|\ph\|\|\psi\|.
\end{split}
\end{equation}
Therefore (iii) implies (ii).

Next we assume (ii). Then $TD(A)\subset D(A)$ and
$[A,T]:D(A)\subset\H\to\H$ has a unique extension to a bounded
operator $\ad{A}{T}$ on $\H$. The mapping
$$
   \tau\mapsto e^{-iA\tau}\ad{A}{T}e^{iA\tau}\psi
$$
is continuous, and hence \eqref{c1a-1} implies that
\begin{equation}\label{c1a-3}
  e^{-iAs}Te^{iAs}\psi - T\psi = -i\int_0^s
   e^{-iA\tau}\ad{A}{T}e^{iA\tau}\psi\,d\tau
\end{equation}
for each $\psi\in \H$. Since the r.h.s is continuously
differentiable in $s$, so is the l.h.s, and thus $T\in
C^{1}(A)$.
\end{proof}

Let $A_s=(e^{iAs}-1)/is$, which is a bounded approximation of $A$. Then
\begin{equation}\label{Ws-As}
  \frac{1}{s}\left(e^{-iAs}Te^{iAs}-T\right) = -i e^{-iAs}\ad{A_s}{T}.
\end{equation}
Hence, by Proposition~\ref{C1-tool}, a bounded operator $T$ is of class $C^1(A)$ if and only if
$ \liminf_{s\to 0+}\|\ad{A_s}{T}\|<\infty$. The following proposition gives an
analogous characterization of the class $C^2(A)$.

%%%%%%%%%%%%%%%%%%%  C2- Tool  %%%%%%%%%%%%%%%%%%%%%%%%%%%%%%%%%%%%%%%

\begin{prop}\label{C2-tool}
Let $A=A^*: D(A)\subset\H\to\H$ and let $T$ be a bounded operator of class
$C^1(A)$. Then $T$ is of class $C^2(A)$ if and only if
\begin{equation}
  \label{C2-cond}
  \liminf_{s\to 0+}\|\add{A_s}{T}\|<\infty.
\end{equation}
\end{prop}

\emph{Remark.} This is a special case of \cite[Lemma
6.2.3]{Amreinetal1996} on the class $C^k(A)$. We include the proof for the
convenience of the reader.

\begin{proof}
Since $T$ is of class $C^1(A)$ the commutator $[A,T]$ extends to a
bounded operator $\ad{A}{T}$ on $\H$ and
\begin{equation}\label{c2-1}
  i\frac{d}{ds} e^{-iAs}T e^{iAs}\ph = e^{-iAs}\ad{A}{T}e^{iAs}\ph
\end{equation}
for all $\ph\in \H$. By Proposition~\ref{C1-tool} the right hand
side is continuously differentiable if and only if
\begin{equation}\label{c2-2}
  |\sprod{A\ph}{\ad{A}{T}\psi} - \sprod{\ph}{\ad{A}{T}A\psi}|
  \leq c\|\ph\| \|\psi\|, \qquad \text{for}\ \ph,\psi\in D(A)
\end{equation}
with some finite constant $c$. To prove that \eqref{c2-2} is equivalent to \eqref{C2-cond}, it is useful to
introduce the homomorphism $W(s):T\mapsto e^{-iAs}Te^{iAs}$ on the algebra
of bounded operators. By \eqref{c1a-3}
$$
   (W(s)-1)T = -i\int_0^s \rd \tau_1 W(\tau_1)\ad{A}{T}
$$
and therefore
\begin{eqnarray}
  \frac{1}{s^2}(W(s)-1)^2T &=&
  \frac{-i}{s^2}\int_0^s\rd\tau_1 (W(s)-1)W(\tau_1)\ad{A}{T}\nonumber \\
  &=& \frac{-1}{s^2}\int_0^s\rd\tau_1\int_0^s\rd\tau_2 W(\tau_1+\tau_2)[A,\ad{A}{T}]\label{c2-3}
\end{eqnarray}
in the sense of quadratic forms on $D(A)$, that is,
$$
  \sprod{\ph}{W(\tau_1+\tau_2)[A,\ad{A}{T}]\psi}:=\sprod{A\ph}{W(\tau_1+\tau_2)\ad{A}{T}\psi}-
   \sprod{\ph}{W(\tau_1+\tau_2)\ad{A}{T}A\psi}
$$
for  $\ph,\psi\in D(A)$. Since the right hand side is continuous
as a function of $\tau_1+\tau_2$, it follows from \eqref{c2-3}, as
in the proof of Proposition~\ref{C1-tool}, that
\begin{eqnarray*}
  |\sprod{A\ph}{\ad{A}{T}\psi} - \sprod{\ph}{\ad{A}{T}A\psi}| &=&
  \lim_{s\to 0+} \frac{1}{s^2}|\sprod{\ph}{(W(s)-1)^2T\psi}| \\
  &\leq & \liminf_{s\to 0+} \frac{1}{s^2}\|(W(s)-1)^2T\|\|\ph\|\|\psi\|.
\end{eqnarray*}
Since, by \eqref{Ws-As},
$$
   \frac{1}{s^2}(W(s)-1)^2T = -e^{-2iAs}\mathrm{ad}_{A_s}^2(T),
$$
condition \eqref{C2-cond} implies \eqref{c2-2}. Conversely, by \eqref{c2-3} condition
\eqref{c2-2} implies that $s^{-2}\|(W(s)-1)^2T\|\leq c$
for all $s>0$, which proves \eqref{C2-cond}.
\end{proof}

%%%%%%%%%%%%%%%%%%%%%%%%%%%%%%   commutator-tool  %%%%%%%%%%%%%%%%%%%%%

\begin{lemma}\label{lm:com-H-A}
Suppose that $H$ is locally of class $C^1(A)$ in $\Omega\subset\R$
and that $e^{iAs}D(H)\subset D(H)$ for all $s\in \R$. Then, for
all $f\in C_0^{\infty}(\Omega)$ and all $\ph\in\H$
$$
   f(H)[H,iA]f(H)\ph = \lim_{s\to
   0}f(H)\left[H,\frac{e^{iAs}-1}{s}\right]f(H)\ph.
$$
\end{lemma}

\begin{proof}
By Equation 2.2 of \cite{Sahbani1997},
\begin{equation}\label{leibnitz-com}
f(H)[H,iA]f(H) = [Hf^2(H),iA] - Hf(H)[f(H),iA] - [f(H),iA] Hf(H),
\end{equation}
where, by assumption, $f(H)$ and $Hf^2(H)$ are of class $C^1(A)$.
Since, by \eqref{Ws-As}
$$
   [T,iA]\ph = -i\lim_{s\to 0}\ad{A_s}{T}\ph
$$
for every bounded operator $T$ of class $C^1(A)$, it follows from
\eqref{leibnitz-com}, the Leibniz-rule for $\mathrm{ad}_{A_s}$ and
the domain assumption $A_sD(H)\subset D(H)$, that
\begin{eqnarray*}
  \lefteqn{f(H)[H,iA]f(H)\ph}\\
   &=& -i\lim_{s\to 0}\Big(\ad{A_s}{Hf^2(H)}
  -Hf(H)\ad{A_s}{f(H)} - \ad{A_s}{f(H)}Hf(H)\Big)\ph\\
  & = &-i\lim_{s\to 0}f(H)\ad{A_s}{H} f(H)\ph.
\end{eqnarray*}
\end{proof}

\end{document}